\documentclass[aps, pre, longbibliography, superscriptaddress, tightenlines, nofootinbib, amsfonts, amsmath, amssymb, floatfix, twocolumn, showpacs]{revtex4-1}  

\usepackage{graphics}
\usepackage{graphicx}
\usepackage{amsmath}
\usepackage{amssymb}
\usepackage[usenames,dvipsnames,svgnames,table]{xcolor}
\usepackage{times}
\usepackage{comment}
\usepackage{tikz}       
\usepackage[smallscripts]{moresize} 

\definecolor{darkgreen}{rgb}{0,0.5,0}
\definecolor{darkblue}{rgb}{0,0,0.9}

\usepackage[colorlinks=true,linkcolor=darkblue,citecolor=darkblue,urlcolor=darkblue]{hyperref}

\DeclareMathAlphabet{\mathpzc}{OT1}{pzc}{m}{it}

\DeclareFontFamily{OT1}{pzc}{}
\DeclareFontShape{OT1}{pzc}{m}{it}{<-> s * [1.300] pzcmi7t}{}
\DeclareMathAlphabet{\mathpzc}{OT1}{pzc}{m}{it}

\DeclareMathSymbol{\shortminus}{\mathbin}{AMSa}{"39}

\newcommand{\fett}[1]{\boldsymbol{#1}}
\newcommand{\dd}{{\rm{d}}}
\newcommand{\ii}{{\rm{i}}}
\newcommand{\be}{\begin{equation}}
\newcommand{\ee}{\end{equation}}

\newcommand{\sdfrac}[2]{\mbox{\small$\displaystyle\frac{#1}{#2}$}}
\newcommand{\ws}{\hspace{0.03cm}}
\newcommand{\PN}{\text{P}_{\!\text{\tiny $N$}}}
\newcommand{\PNof}[1]{\text{P}_{\!\text{\tiny $#1$}}}
\newcommand{\PKG}{\text{P}_{\!\text{\ssmall $K$}_{\!\ws\text{\tiny G}}}}
\newcommand{\KG}{K_{\text{\ssmall G}}}
\newcommand{\PKP}{\text{P}_{\!\text{\ssmall $K$}_{\!\ws\text{\tiny P}}}}
\newcommand{\KP}{K_{\text{\ssmall P}}}
\newcommand{\ts}{\hspace{0.001cm}}
\newcommand{\mtau}{\!\ws\mathpzc{t}\ts}
\newcommand{\mtstar}{\!\ws\mathpzc{t}_{\hspace{0.02cm}\star}\ts}
\newcommand{\mx}{\mathpzc{x}}
\newcommand{\mv}[1]{\mathpzc{v}_{\{\ws\! #1 \ws\!\}}}
\newcommand{\mvini}{\mathpzc{v}_0}
\newcommand{\uN}{u_{\!\ws\text{\tiny $N$}}}
\newcommand{\cN}{c_{\!\ws\text{\tiny $N$}}}
\newcommand{\calcN}{\mathpzc{c}_{\!\ws\text{\tiny $N$}}}
\newcommand{\mT}[1]{{\cal T}(\text{\fontsize{8}{8}\selectfont $#1$})}
\newcommand{\mE}[1]{\delta \!\hspace{0.02cm} E(\text{\fontsize{8}{8}\selectfont $#1$})}
\newcommand{\xstar}{x_{\!\!\hspace{0.08cm}\star}}
\newcommand{\mminus}{\fett{\shortminus}}
\newcommand{\uPlatz}{u_{\text{\tiny P}}}

\definecolor{lime}{HTML}{A6CE39}
\DeclareRobustCommand{\orcidicon}{
	\begin{tikzpicture}
	\draw[lime, fill=lime] (0,0) 
	circle [radius=0.14] 
	node[white] {{\fontfamily{qag}\selectfont \tiny ID}};
	\draw[white, fill=white] (-0.0625,0.095) 
	circle [radius=0.007];
	\end{tikzpicture}
	\hspace{-2mm}
}

\foreach \x in {A, ..., Z}{%
	\expandafter\xdef\csname orcid\x\endcsname{\noexpand\href{https://orcid.org/\csname orcidauthor\x\endcsname}{\noexpand\orcidicon}}
}

\newcommand{\mnras}{Mon. Not. Roy. Astron. Soc.}

\newcommand{\physrep} {Phys. Rep.}

\newcommand{\jcap}{J. Cosmol. Astropart. Phys.}
\newcommand{\prf}{Phys. Rev. Fluids}
\newcommand{\aap}{Astron. Astrophys.}
\newcommand{\siam}{SIAM J. Numer. Anal.} 
\newcommand{\jcomput}{J. Comput. Phys.}    
\newcommand{\jfm}{J. Fluid Mech.}  

\begin{document}

\title[Singularities in Burgers]{Eye of the Tyger: early-time resonances  and singularities  in the inviscid Burgers equation}

\author{Cornelius Rampf${\,{\tiny\orcidA{}}}$\phantom{II}}\email{cornelius.rampf@univie.ac.at}
\affiliation{Department of Mathematics, University of Vienna, Oskar-Morgenstern-Platz 1, 1090 Vienna, Austria}
\affiliation{Department of Astrophysics, University of Vienna, T\"urkenschanzstraße 17, 1180 Vienna, Austria}

\author{Uriel Frisch${\,{\tiny\orcidB{}}}$\phantom{II}}\email{uriel@oca.eu}
\affiliation{Laboratoire Lagrange, Universit\'e C\^ote d'Azur, Observatoire de la C\^ote d'Azur, CNRS, Blvd de l'Observatoire, CS 34229, 06304 Nice, France}

\author{Oliver Hahn${\,{\tiny\orcidC{}}}$\phantom{II}}\email{oliver.hahn@univie.ac.at}
\affiliation{Department of Mathematics, University of Vienna, Oskar-Morgenstern-Platz 1, 1090 Vienna, Austria}
\affiliation{Department of Astrophysics, University of Vienna, T\"urkenschanzstraße 17, 1180 Vienna, Austria}

\date{\today}

\begin{abstract}
We chart a singular landscape in the temporal domain of the inviscid Burgers equation in one space dimension for sine-wave initial conditions. These so far undetected complex singularities are arranged in an eye shape centered around the origin in time. Interestingly, since the eye is squashed along the imaginary time axis, complex-time singularities can become physically relevant at times well before the first real singularity---the pre-shock.  Indeed, employing a time-Taylor representation for the velocity around $t=0$, loss of convergence occurs roughly at 2/3 of the pre-shock time for the considered single- and multi-mode models. Furthermore, the loss of convergence is accompanied by the appearance of initially localized resonant behaviour which, as we claim, is a temporal manifestation of the so-called tyger phenomenon, reported in Galerkin-truncated implementations of inviscid fluids [S.\,S.\,Ray {\it et al.}, Phys.\ Rev.\ E {\bf 84}, 016301 (2011)]. We support our findings of early-time tygers with two complementary and independent means, namely by an asymptotic analysis of the time-Taylor series for the velocity, as well as by a  novel singularity theory that employs Lagrangian coordinates. 

Finally, we apply two methods that reduce the amplitude of early-time tygers. One is tyger purging which removes large Fourier modes from the velocity, and is a variant of a procedure known in the literature. The other method realizes an {\it iterative} UV completion, which, most interestingly,  iteratively restores the conservation of energy once the Taylor series for the velocity diverges. Our techniques are straightforwardly adapted to higher dimensions and/or applied to other equations of hydrodynamics.

\end{abstract}


\maketitle

\section{Introduction and basic formulation}

The Burgers equation is of interest in many scientific disciplines \cite{BEC20071}, ranging from ultra-large, cosmological scales serving as a reduced model for cosmic structure formation \cite{1994A&A...289..325V,Bernardeau2002}, to small scales such as applied to many body-chaos in condensed matter problems \cite{PhysRevLett.127.124501}.
The inviscid case has a well-known solution in one space dimension obtained through the method of characteristics, which is valid until the appearance of the first real singularity, the pre-shock, when the gradient of the velocity becomes singular. It is perhaps because of this exact solution that the temporal regime until pre-shock is frequently considered as being of little interest.

However, for many related considerations, such as the blow-up problem in incompressible Euler flow or Navier--Stokes, it is precisely this temporal regime that needs to be resolved to very high precision. Furthermore, many state-of-the-art numerical techniques employ an Eulerian specification of the flow field, where the method of characteristics is often not constructive.
In the present paper we highlight a path where, even in an Eulerian setup, the method of characteristics can be used to gain precious information about the singular structure of the underlying equation.

We focus on the inviscid one-dimensional (1D) Burgers equation
\be \label{eq:burger}
 \partial_t u + u \ws \partial_x u  = 0 \,,   \qquad \quad u(x,0) = u_0(x) \,,
\ee
and analyze the emergence of non-analyticity, starting from  smooth and analytic initial conditions. For simplicity we limit ourselves to $2\pi$-periodic initial data.

One way to investigate the analytic structure of~\eqref{eq:burger} is to consider  a time-Taylor series representation for the velocity,
\be \label{eq:Ansatz}
   u(x,t) = \sum_{n=0}^\infty u_n(x) \ws t^n \,,
\ee
where the $u_n$  are time-Taylor coefficients that are easy to determine (see below).
The range of validity of the time-Taylor series~\eqref{eq:Ansatz} is determined by the radius of convergence~$R$, which is the  distance between the expansion point and the closest singularity(ies) in the complex-time plane. It is an essential aspect in the present paper to assess the singular behaviour of the velocity. 
We will see that the series loses convergence at the time of first pre-shock, denoted with $t_\star$, which is the instant when $\partial_x u$ becomes singular (see e.g.\  \cite{10.1007/3-540-10694-4_44}). This should not bother us, as  we are here interested in the analysis of singular behaviour that occurs at times before~$t_\star$.

 Plugging~\eqref{eq:Ansatz} into Burgers' equation~\eqref{eq:burger} and identifying the coefficients of the involved powers in $t$, one obtains the simple  recursive relation ($n \geq0$)
\be \label{eq:rec}
  u_{n+1} =   \tfrac{-1}{n+1} \sum_{i+j=n} u_i \ws \partial_x u_j \,,
\ee
which, of course, only requires the initial data~$u_0$ as input.
For general applications, the Taylor series for the velocity is truncated which, as we elucidate in the following, comes with important consequences 
for triggering resonant behaviour dubbed ``tygers'' (named after William Blake's poem).
Let us define the projection operator~$\PN$ associated with the truncation order $N$, such that only Taylor coefficients  until 
$N$ are retained, i.e.,
\be \label{eq:PNu}
   \PN u = \sum_{n=0}^N u_n t^n \,.
\ee
For initial data with a finite number of modes, the truncated velocity  $\PN u$ contains only a limited number of non-zero Fourier modes (see below). Hence, $\PN u$ is bandlimited in Fourier space and,  in some sense, the operator $\PN$ acts as an effective Galerkin projector, commonly denoted with~$\PKG$.

This paper is organized as follows. General properties of the time-Taylor solutions, as well as the birth of early-time tygers are discussed in Sec.\,\ref{sec:anal-BETT}, on the basis of a simple single-mode model.
Sections~\ref{sec:asy-euler} and~\ref{sec:asy-lagrange} provide respectively an asymptotic analysis in Eulerian space, and a singularity theory in Lagrangian space.
Then, in Sec.\,\ref{sec:arrestETTs}, we provide two means for halting early-time tygers, namely through tyger purging and an iterative UV completion. 
In Sec.\,\ref{sec:multi}, we generalize the asymptotic analysis and singularity theory to a model with two-mode initial data (and beyond). We conclude in Sec.\,\ref{sec:sum-out} where we also provide an outlook.

\section{Phenomenology and asymptotic analysis}\label{sec:pheno-asy}

\subsection{Analytic solutions and the birth of early-time tygers} \label{sec:anal-BETT}

Let us begin with a simple single-mode case  with periodic initial data
\be \label{eq:ICsinglemode}
  u_0 = - \sin x \,,
\ee
for which the first pre-shock occurs at $t= t_\star =1$  at location $x = 0$.  
Using the recursive relation~\eqref{eq:rec}, one easily finds
\begin{subequations} \label{eqs:uns}
\begin{align}
  u_1 &= -  \sdfrac{1}{2} \sin (2x) \,, \\
  u_2 &=  \sdfrac{1}{8} [\sin x  - 3 \sin(3 x)]  \,, \\
  u_3 &=  \sdfrac 1 6 [ \sin(2 x) - 2 \sin(4 x)] \,, \\
  u_4 &= \sdfrac{-1}{384} [  2 \sin x - 81 \sin(3 x)  + 125 \sin(5 x)] \,, \\
   &\,\,\, \rotatebox{90}{$\cdots$}  \nonumber \\
  \uN &= \cdots + \cN \sin[(N+1)x] \,, \label{eq:un}
\end{align}
\end{subequations}
where $\cN$ is a constant coefficient. For later applications, we have explicitly determined the first 70 Taylor coefficients.
Due to the spatial structure of the time-Taylor coefficients $\uN$, it easily follows that the Taylor-truncated velocity $\PN u$ has bounded support in Fourier space. For the present choice of initial data, the largest wavenumber is $k= \pm( N+1)$.

Actually, in the limited case of the single-wave initial data~\eqref{eq:ICsinglemode}, there exists an exact analytical solution for the velocity based on work by George W.\ Platzman \cite{1964Tell...16..422P} ($0 \leq t < t_\star$)
\be \label{eq:uPlatz}
  u(x,t) = - 2 \sum_{k=1}^\infty \frac{J_k(k t)}{kt} \sin(k x)\,, 
\ee
where $J_k$ is the Bessel function of the first kind. We will comment on this solution further below, but for the moment note that Eq.\,\eqref{eq:uPlatz} can be used to determine very efficiently the time-Taylor coefficients $u_n$ of~\eqref{eq:Ansatz}, simply by expanding the r.h.s.\ of~Eq.\,\eqref{eq:uPlatz} around $t=0$. Beyond the single-sine-wave model, however, we are not aware of exact Eulerian solutions.

\begin{figure}[t]
 \centering
   \includegraphics[width=0.9\columnwidth]{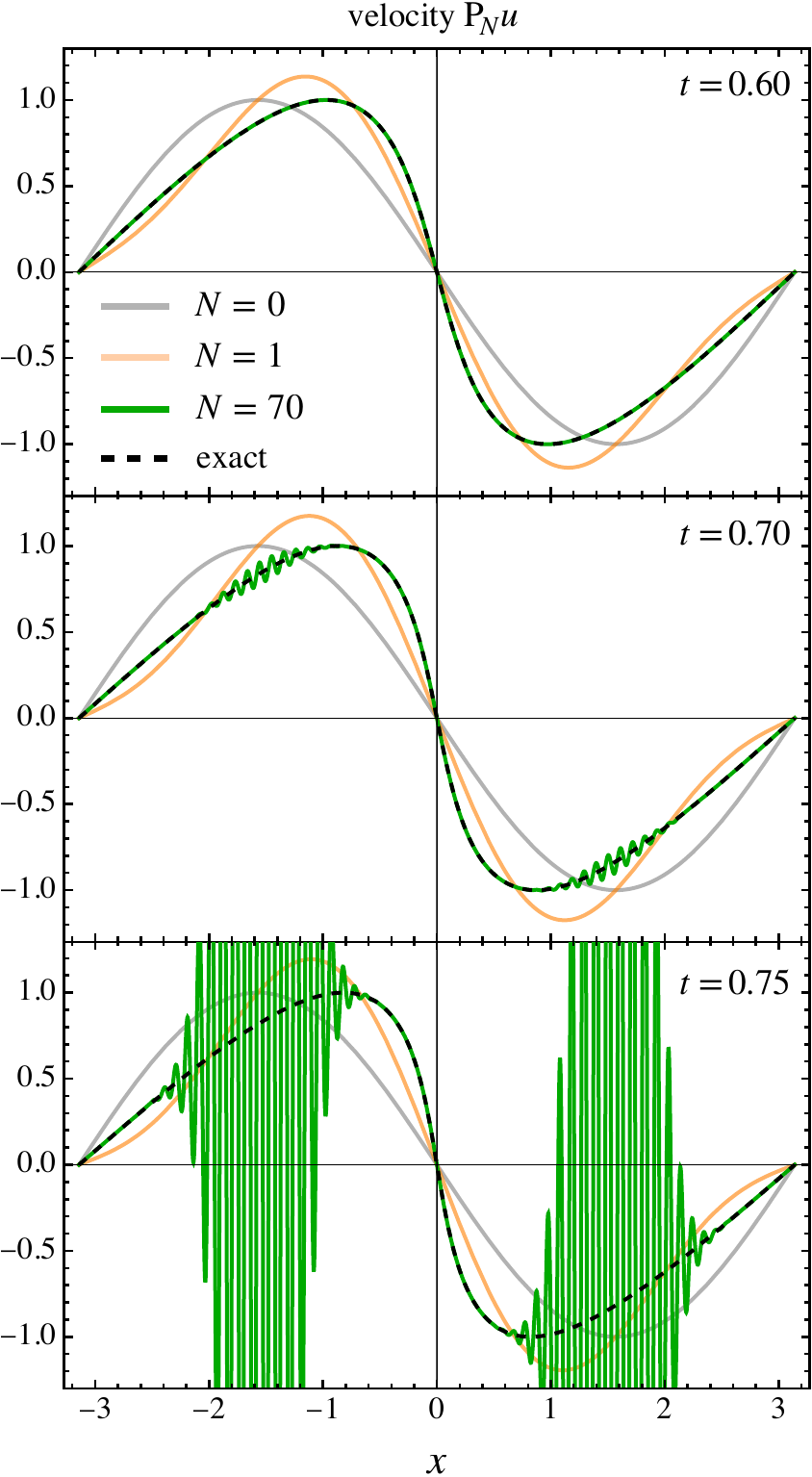}
   \caption{Taylor-series representation of the velocity~\eqref{eq:PNu} for the truncation orders~$N\!=\!0,1,70$ (shown respectively in grey, orange and green), compared against the exact solution (black dashed lines; based on Eq.\,\ref{eq:ZAmap}). 
At early times such as $t=0.6$ (top panel), the Taylor series exhibits converging behaviour with an exponential decay of truncation errors.  However, shortly after ($t=0.7$, central panel), two tygers appear centred at locations $x \simeq \pm \pi/2$. After the birth of the tygers, their amplitude and spatial width is rapidly growing (bottom panel, $t=0.75$), indicating the loss of convergence at times well before the pre-shock ($t_\star=1$).
Location and time of appearance of these tygers are different from those observed recently in numerical setups of the inviscid Burgers equation \cite{2011PhRvE..84a6301R,PhysRevE.87.033017,2018PhRvF...3a4603C,2017RSPSA,2020PhRvR...2c3202M}. 
}  
 \label{fig:phasespace}
\end{figure}

In Fig.\,\ref{fig:phasespace} we compare the truncated velocity $\PN u$ for  $N=0,1,70$  against the exact solution at various times. Specifically, the top panel is evaluated at $t=0.6$, where $\PNof{70} u$ appears to agree extremely well with the exact solution: a closer inspection reveals that, starting at $N \gtrsim 30$, the truncation error of $\PN u$ decays exponentially, and is of the order of $10^{-4}$ at $N=30$ and $10^{-6}$ at $N=70$ for $t=0.6$ (not shown). The exponential decay of the truncation error indicates that the series representation~\eqref{eq:Ansatz} of the velocity is converging at such early times; see section~\ref{sec:asy-euler} for further details on convergence.

However, as shown in the central and bottom panels of Fig.\,\ref{fig:phasespace}, convergence is lost at later times, with the appearance of two resonances centred  around $x = \pm \pi/2$ 
(see Fig.\,\ref{fig:Rthetanu} and accompanying text for identifying the precise location of the tygers). We claim that these features are certainly related to the tyger phenomenon as recently analyzed \cite{2011PhRvE..84a6301R,PhysRevE.87.033017,2018PhRvF...3a4603C,2017RSPSA,2020PhRvR...2c3202M}. However there are significant differences: First of all, for the present choice of single-mode initial conditions, the commonly known tyger phenomenon would appear only at locations with positive space derivative and matched velocity at the pre-shock location. Thus, since the pre-shock velocity is zero,  Refs.~\cite{2011PhRvE..84a6301R,PhysRevE.87.033017,2018PhRvF...3a4603C,2017RSPSA,2020PhRvR...2c3202M} would observe a single tyger appearing around the location $x \simeq -\pi$.
Second, in the numerical setups of  Refs.~\cite{2011PhRvE..84a6301R,PhysRevE.87.033017,2018PhRvF...3a4603C,2017RSPSA,2020PhRvR...2c3202M}, the tyger would appear at times much closer to the pre-shock time $t_\star =1$, specifically at a time depending on the used Galerkin truncation of the simulation (which is typically much larger than the considered Taylor truncations of the velocity).

In the following section, we analyze why these early-time-tygers are born, and why they appear  at the shown locations.
Before doing so, we remark that Platzman's exact result for the velocity (Eq.\,\ref{eq:uPlatz}) does not display the birth of tygers. The reason for that lies in the derivation of his result \cite{1964Tell...16..422P}, which can be sketched as follows. The starting point is the Fourier representation of the velocity $u(x,t) =\sum_{k=1}^\infty  \uPlatz (k,t) \sin kx$ in a sine-wave basis, with Fourier coefficient 
$\uPlatz(k,t) = (1/\pi) \int_{-\pi}^{+\pi} u(x,t) \sin(k x)\, \dd x$. Integrating the r.h.s.\ of $\uPlatz(k,t)$ by parts and substituting the {\it exact Lagrangian-coordinates solution} (eq.\,\ref{eq:ZAmap}), Platzman showed that the $k$th Fourier coefficient is $\uPlatz(k,t) = - 2 J_k(kt)/(kt)$ upon identification, leading precisely to Eq.\,\eqref{eq:uPlatz}. Thus, Platzman's solution exploits the Lagrangian-coordinates solution which is 
time analytic within the real-valued domain $0 \leq t < t_\star$ and thus, similarly as with the Lagrangian solution,  no tygers do appear.

\subsection{Asymptotic analysis in Eulerian coordinates} \label{sec:asy-euler}

Figure~\ref{fig:phasespace} suggests that the convergence of
\be \label{eq:uTaylorRep}
  u = \sum_{n=0}^\infty u_n \ws t^n
\ee
is lost, at some time between $t = 0.6$ and~$0.7$. Furthermore, it is also clear that questions about convergence of~\eqref{eq:uTaylorRep} are not only a matter of time but also of space. Indeed, convergence is, at first, lost in a narrow region centered around $x = \pm\pi/2$, but eventually spreads out to a much wider range of spatial scales.

It is thus instructive to first search for the space-dependent radius of convergence $R(x)$, given by the Cauchy--Hadamard formula
\be \label{eq:CH}
  \frac{1}{R(x)} = \limsup_{n \to \infty}  \sqrt[n]{|u_n(x)|}   \,, \qquad R>0\,,
\ee
for which~\eqref{eq:uTaylorRep} defines an absolutely convergent series over \mbox{$0 \leq t < R(x)$.} Then, it might be natural to define the actual radius of convergence $R_{\rm inf}$ of~\eqref{eq:uTaylorRep} by taking the infimum of  $R(x)$ over all values $x$, i.e., $R_{\rm inf} := \inf_x R(x)$; see e.g.\ \cite{2016JCoPh.306..320P} for similar investigations related to incompressible Euler flow.

Given a finite number of Taylor coefficients, how can we practically estimate the radius of convergence?
One classical way for addressing this question could be the  Domb--Sykes method \cite{1957RSPSA.240..214D}, which graphically exploits the ratio test $1/R = \lim_{n\to \infty}u_n/u_{n-1}$, simply by drawing subsequent ratios of Taylor coefficients against $1/n$, followed by a linear extrapolation to the $y$-intercept. See e.g.\ \cite{2019MNRAS.484.5223R,2021MNRAS.501L..71R} for applications of the Domb--Sykes method within a cosmological context.

We have tested the Domb--Sykes method on~\eqref{eq:uTaylorRep}, however we found that the ratio $u_{n}/u_{n-1}$ swaps signs at consecutive higher orders $n$, and thus, the involved limit $n\to\infty$ in the Domb--Sykes method does not converge. One plausible reason for this non-convergence is that the convergence-limiting singularities are located at complex locations in time, which is accompanied by a non-trivial pattern of signs of the Taylor coefficients. Mercer and Roberts \cite{MercerRoberts1990} generalized the Domb--Sykes method to allow for a pair of complex conjugated singularities, and have applied their method to the study of Poiseuille flow.
This generalized extrapolation method is in principle applicable to any real-valued function with non-trivial sign patterns in its Taylor coefficients; see e.g.\ \cite{2019PhRvD..99k4510G} where the method is applied to the study of phase transitions in finite-temperature quantum chromodynamics.
In the following we utilize (and extend) the Mercer--Roberts method to the inviscid Burgers equations which, to our knowledge, has not yet been performed in the literature.

To proceed, we require some elementary tools from complex analysis. In particular, in what follows it is useful to formally complexify the temporal variable, which we denote with $\mtau$, while it is sufficient to keep the space variable real-valued (since $x$ appears in Eq.\,\ref{eq:uTaylorRep} merely as a parameter).  

Next we assume that the large-$n$ asymptotic behaviour of~\eqref{eq:uTaylorRep} is described  by a  model function $\mathfrak{u}(\mtau)$ that depends on the complexified time $\mtau$, and is built up by an additive pair of complex-conjugated singularities located at $\mtstar$ and $\overline\mtau_{\!\!\hspace{0.07cm}\star}$, i.e., 
\be \label{eq:model}
  \mathfrak{u}(\mtau) = \left(1 - \frac{\mtau}{\mtstar} \right)^{\!\nu} + \left(1 - \frac{\mtau}{\overline\mtau_{\!\!\hspace{0.07cm}\star}} \right)^{\!\nu} \! , \,\,\quad \mtstar := R\ws {\rm e}^{\ii \theta} . 
\ee
Here, $\overline\mtau_{\!\!\hspace{0.07cm}\star}$ denotes the complex conjugate of $\mtstar$,   $\nu$ is a real-valued singularity exponent which is neither zero nor a positive integer, $\theta$  is the phase of the singularity and,  as for $R$, these unknowns are assumed to depend on the real-valued position $x$.
For $|\mtau| < R$, the model function $\mathfrak{u}$ has the following Taylor expansion around $\mtau =0$,
\be \label{eq:modelexpanded}
  \mathfrak{u}(\mtau)  = \sum_{n=0}^\infty 2 (-1)^n \begin{pmatrix} \nu \\ n \end{pmatrix} R^{-n} \cos(n\theta)\,  \mtau^n \,.
\ee
Mercer and Roberts showed that the following estimators can be used for determining the unknowns in the model function~\eqref{eq:model}, and how to relate them to the Taylor coefficients of a standard Taylor series  $u(\mtau) = \sum_{n=0}^\infty  u_n \mtau^n$,
\begin{align}
  B_n^2 &=  \frac{u_{n+1} u_{n-1} - u_n^2}{u_{n} u_{n-2} - u_{n-1}^2} \,, \label{eq:Bnsq} \\ 
 \cos \theta_n &= \sdfrac 1 2 \left( \sdfrac{u_{n-1}}{u_n} B_n   + \sdfrac{u_{n+1}}{u_n } B_n^{-1} \right) \,. \label{eq:costheta} 
\end{align}
Substituting the Taylor coefficients of~\eqref{eq:modelexpanded} into~\eqref{eq:Bnsq} 
and~\eqref{eq:costheta}, one finds respectively for $n\to \infty$ the Mercer--Roberts (MR) estimators \cite{MercerRoberts1990}
\begin{subequations} \label{eqs:MRstandardextra}
\begin{align}
  B_n  &= \sdfrac{1}{R} \left(1  - (\nu +1) \frac 1 n \right) \nonumber \\
   &\quad \times\! \left[ 1 + \sdfrac{\nu+1}{2} \sdfrac{\sin(2n-1)\theta}{\sin \theta}  \frac{1}{n^2} + O(n^{-3}) \right] \!,\! \label{eq:Bn} \\
  \cos \theta_n &= \cos \theta \bigg(   1+ (\nu+1) \left[  1 - \sdfrac{\cos(2n-1)\theta}{\cos \theta} \right] \frac {1}{n^2} \nonumber \\
  &\qquad \qquad + O(n^{-3})  \bigg) . \label{eq:costhetan}
\end{align}
\end{subequations}
By drawing~\eqref{eq:Bn} and \eqref{eq:costhetan} respectively against $1/n$ and $1/n^2$, one can estimate all unknowns in the model function~$\mathfrak{u}(\mtau)$, and hence deduct the leading-order asymptotics of the velocity~\eqref{eq:uTaylorRep}.

Here we must note that the above procedure works well for most parts of the spatial domain of interest, except for locations where the associated phase is close to zero or close to $\pm\pi$, i.e., when the pair of complex singularities is close to the real time axis. Indeed, for such $\theta$'s, the estimators~\eqref{eqs:MRstandardextra}  become unreliable, as terms with higher orders in $1/n$ become large. To handle this drawback, first noted in \cite{MercerRoberts1990}, we 
adapt for all spatial locations with corresponding $|\theta| < 0.1$ (or $|\theta \pm \pi| <0.1$) a computationally demanding non-linear extrapolation method based on the full, un-expanded, form of~$B_n$, as given in~\eqref{eq:Bnsq}.
We find that this non-linear extrapolation method delivers more accurate results than the original Mercer--Roberts method based on Eqs.\,\eqref{eqs:MRstandardextra}.  
Still, as we show later when comparing the results against our theory, accuracy in the extrapolation is in general lost  when the singularities are close to the real time axis.

\begin{figure}[ht]
 \centering
   \includegraphics[width=0.97\columnwidth]{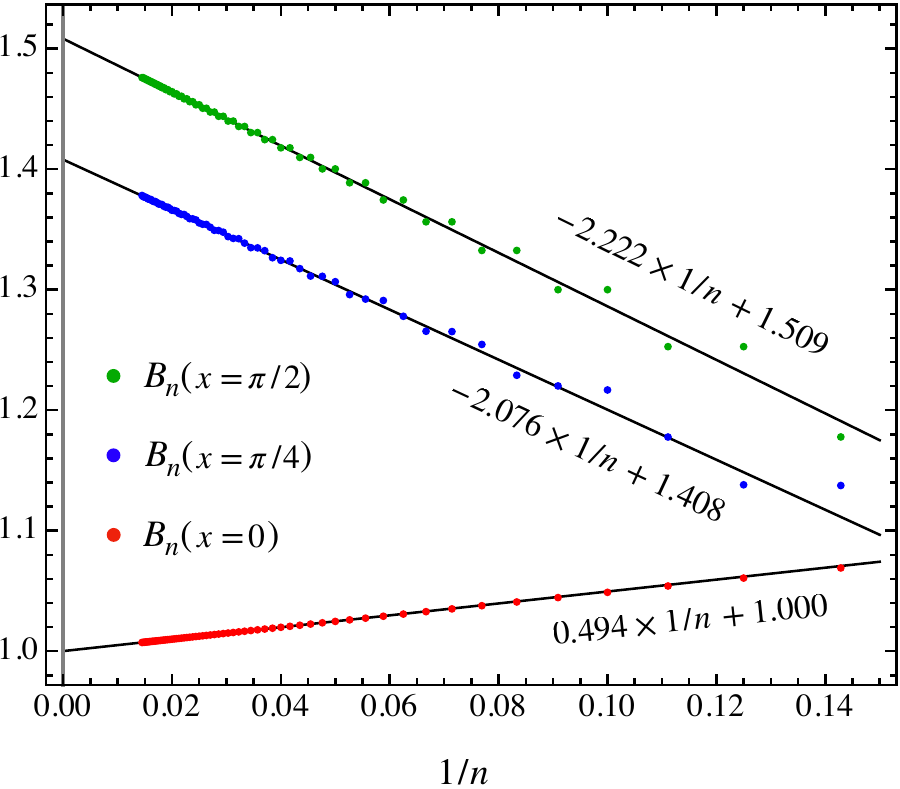}
   \caption{Mercer--Roberts (MR) plot based on the estimator $B_n$ (Eq.\,\ref{eq:Bnsq}) for the velocity up to truncation order $N=70$. Drawn are the $B_n$'s against $1/n$ for the exemplary locations $x=\pi/2,\pi/4,0$ (respectively shown with green, blue and red dots).
Linear extrapolation to the $y$-intercept then provides an estimate of the inverse radius of convergence, essentially by exploiting relation~\eqref{eq:Bn}. Specifically, at $x=0$ which is the location of the pre-shock, we find $R\simeq 1.000$ agreeing to high precision with $t_\star=1$. The lowest radius of convergence of $1/1.509 \simeq 0.663$ is achieved at the location $x= \pi/2$.
See Fig.\,\ref{fig:Rthetanu} for the full spatial dependence of~$R$.
}   \label{fig:mercer}
\end{figure}

To demonstrate the idea behind the extrapolation method, we show in Fig.\,\ref{fig:mercer} the Mercer--Roberts plot for three exemplary locations $x\!=\!\pi/2,\pi/4,0$ (respectively shown with green, blue and red dots). Specifically, for fixed location $x$, we determine the $B_n$'s as defined in Eq.\,\eqref{eq:Bn} up to truncation order $n=N=70$, using the Taylor coefficients~\eqref{eqs:uns} of~$u$ as the input. 
Drawing then $B_n$ against $1/n$, it is seen that the $B_n$'s settle into a linear behaviour for $n \gtrsim 30$. This justifies the use of a linear extrapolation to the $y$-intercept, from which one can read off the inverse of the radius of convergence, essentially as a consequence of exploiting Eq.\,\eqref{eq:Bn} in the limit~$n \to \infty$.

In Fig.\,\ref{fig:mercer} we have also added the formulas resulting from said linear extrapolation. For $x=0$ which for the present initial conditions marks the position of the pre-shock at time $t= t_\star =1$, we find a $y$-intercept of about $1.000061$ from which it follows that $1/R(\text{\small $x=0$}) =1$ to high accuracy. 
Thus, the estimated radius of convergence at pre-shock location agrees to high precision
with  the time of pre-shock,
indicating that the series representation~\eqref{eq:uTaylorRep} is doomed to fail at that time. As mentioned before, this failure is not unexpected, as $\partial_x u$ becomes singular at the pre-shock location, and a spatial singularity can easily translate into a temporal one. Indeed, the  Burgers equation reads $\partial_t u = - u \partial_x u$ and, based on a Fuchsian argument \cite{Moser1959}, a spatial singularity on the right-hand-side is compensated by a temporal singularity on the left-hand-side.

\begin{figure}
 {\centering
   \includegraphics[width=0.99\columnwidth]{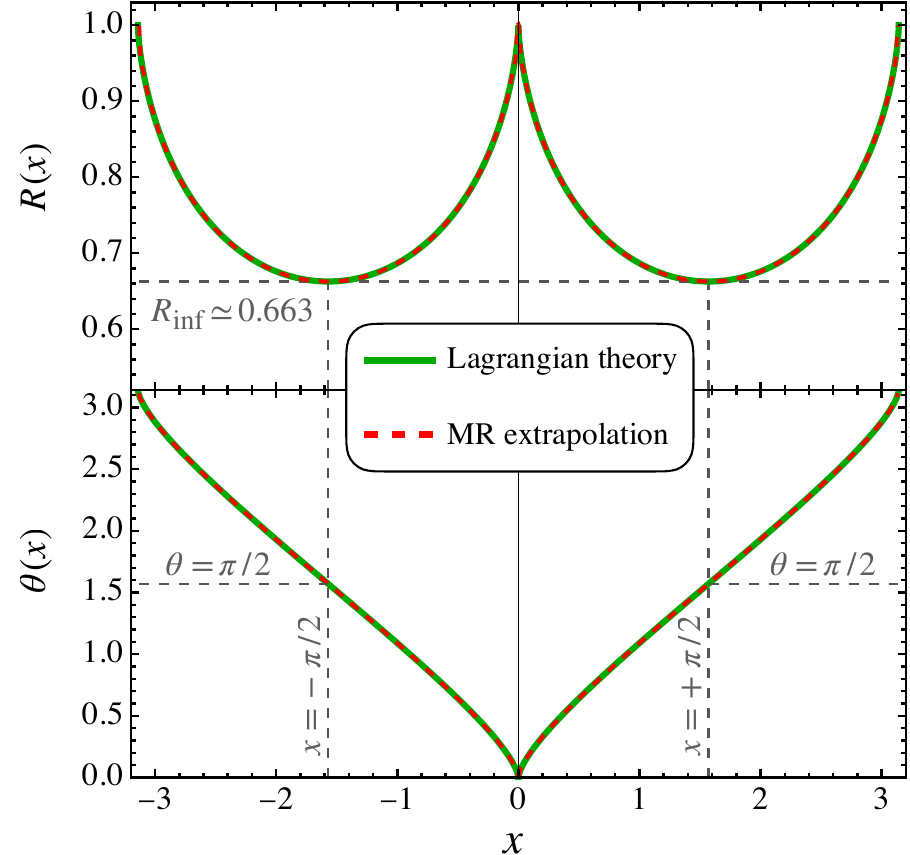}
  }
   \caption{Locations of the convergence-limiting singularities $\mtstar = R \exp(\ii \theta)$ as a function of $x$. The red (dashed) lines are obtained by applying the MR extrapolation technique for all real-valued locations in Eulerian coordinates  (Eqs.\,\ref{eqs:MRstandardextra}), while the green (solid) lines are obtained using the Lagrangian theory (see section~\ref{sec:asy-lagrange}). The minimal radius of convergence of $R_{\rm inf} = 0.663$ is achieved at the locations $x = \pm \pi/2$, which come with a phase of $\theta =\pi/2$. These singularities are thus perfectly aligned along the imaginary time axis and are associated with early-time-tygers. See Fig.\,\ref{fig:Lag2} for the eye of the tyger.}   \label{fig:Rthetanu}
\end{figure}

Interestingly, Fig.\,\ref{fig:mercer} indicates that for $x=\pi/2$ the radius of convergence is $R \simeq 1/1.509$, which appears to imply that~\eqref{eq:uTaylorRep} loses convergence at a time well before the pre-shock, namely at $t \simeq 1/1.509\simeq  0.66272$.
To elucidate this situation in more detail, we show in Fig.\,\ref{fig:Rthetanu} results from the MR extrapolation technique  of the radius of convergence and of the associated phase, now over the whole spatial range (red dashed lines; green lines denote theoretical results, see section~\ref{sec:asy-lagrange}).
At pre-shock location $x=0$, the convergence-limiting singularity has vanishing phase and, as mentioned above, a modulus of unity. Thus, as expected, the pre-shock singularity is of real nature.
At $x =\pm \pi/2$, we obtain numerically that $\theta = \pi/2$ to a precision of nine~significant digits, strongly indicating that the convergence-limiting singularity is exactly aligned along the imaginary axes in time. 
Since the space-dependent radius of convergence takes its minimal value at $x =\pm \pi/2$, we have thus for the actual radius of convergence $R_{\rm inf} = \inf_x R(x) = 0.66272$. 
Looking back at Fig.\,\ref{fig:phasespace}, it becomes evident that the resonant behaviour shown in the middle and bottom panel is driven by complex-time singularities.

\subsection{Singularity theory in Lagrangian frame of reference} \label{sec:asy-lagrange}

The results shown above are of phenomenological nature and rely on an approximative extrapolation technique. Hence, these results do not attempt to address fundamental questions of the origin of the singular structure in the inviscid Burgers equation.
Here we show that these singularities can be fully described by a theory that at its heart employs the method of characteristics (see section~\ref{sec:multi} for the theory applied to multi-mode initial conditions).

For this we employ the direct Lagrangian map $a \mapsto x$ from initial ($t=0$) position $a$ to the current/Eulerian position $x$ at time~$t$. The velocity is defined through the characteristic equation $u (x(a,t),a) = \dot x(a,t)$, where the overdot denotes the Lagrangian (convective) time derivative. 
Employing Lagrangian coordinates, the inviscid Burgers equation~\eqref{eq:burger} reduces to
$\ddot x(a,t) =0$, which has the well-known solution
\be \label{eq:ZAmap}
   x(a,t) = a + t\ws u_0(a) = a - t \sin a
\ee  
 (see e.g.\ \cite{zbMATH03915880,1994A&A...289..325V}). The Jacobian of the transformation 
\be
  J(a,t) := \sdfrac{ \partial x}{\partial a} = 1 - t \cos a 
\ee 
vanishes at pre-shock time $t=t_\star =1$ at location $a =a_\star =0 = \xstar$ (modulo $2\pi$-periodic repetitions).

In section~\ref{sec:asy-euler} we have seen that singularities appear in Eulerian space at times well before $t_\star=1$. To assess this scenario within the present description, we must allow the fluid variables to also take complex values. Thus, we {\it complexify the Lagrangian and Eulerian locations} and denote them respectively with $\mathfrak{a}$ and $\mathpzc{x}$.  Additionally, as in section~\ref{sec:asy-euler}, we employ the complexified time denoted with~$\mtau$.

\begin{figure}[t]
 {\centering
   \includegraphics[width=0.9\columnwidth]{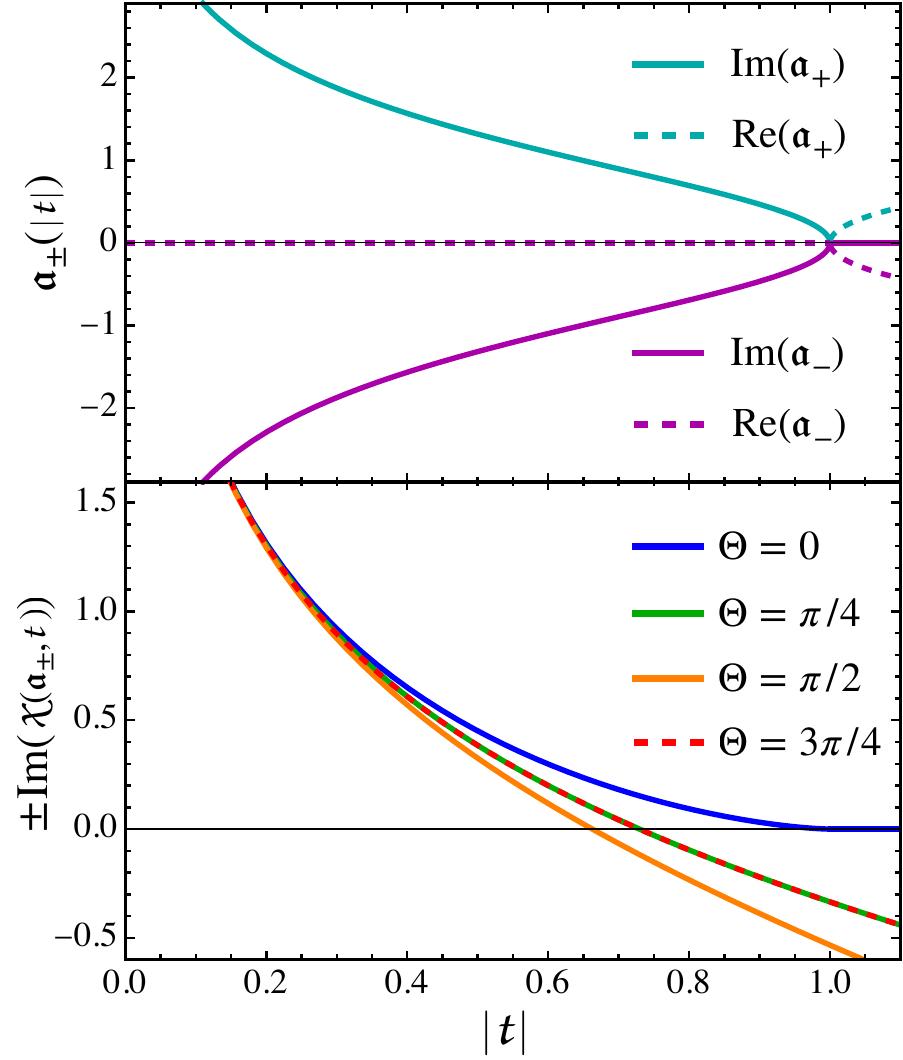}
  }
   \caption{{\it Top panel:} Evolution of complex Lagrangian roots $\mathfrak{a}_\pm$, obtained from solving $J=0$ as a function of~$\mtau=|\mtau| \exp(\ii \Theta)$ with $\Theta=0$. Both roots are purely imaginary for $|\mtau| <t_\star$ and become real precisely at the time of the pre-shock ($t_\star =1$). {\it Bottom panel:}  Evolution of $\pm {\rm Im}(\mx)$ evaluated at the root~$\mathfrak{a}_\pm$, for an exemplary selection of phases $\Theta = 0,\pi/4,\pi/2,3\pi/4$ (respectively shown in blue, green, orange and red-dashed lines).  The vanishing of $\pm {\rm Im}(\mx(\mathfrak{a}_\pm))$ marks the instant~$\mtau =\mtstar = R \exp(\ii \theta)$ when the pre-shock singularity becomes real at the current location in Eulerian space.}   \label{fig:Lag1}
\end{figure}

Now, let us consider complex times $\mtau$ with  $|\mtau| \leq t_\star=1$, and search for the complex Lagrangian roots, dubbed $\mathfrak{a}_\pm$, for which the Jacobian  of the Lagrangian map vanishes, i.e.,
\be
  \mathfrak{a} = \mathfrak{a}_\pm :  \qquad  {\cal J} =  \sdfrac{\partial \mathpzc{x}}{\partial \mathfrak{a}} = 0 \,.
\ee
One easily finds the two exact roots
\begin{align}
  \mathfrak{a}_\pm &= \pm \arccos \left( \frac 1 \mtau \right)  \,, 
\intertext{which imply the current/Eulerian locations}
   \mathpzc{x}(\text{\small $\mathfrak a=\mathfrak{a}_\pm$}, \text{\small $\mtau$}) &= \pm \left[ \arccos\left(\frac 1 \mtau \right) - \mtau \sqrt{1-\frac 1 {\mtau^{2}}} \right] \,. \label{eq:complexX}
\end{align}
In the upper panel of Fig.\,\ref{fig:Lag1}, we show the evolution of the complex roots as a function of $\mtau = |\mtau|$. For $\mtau = |\mtau|<1$, these roots are purely imaginary, but if $\mtau$ is not aligned along the real time axis, the roots are in general complex (not shown). 
Could these complex roots of ${\cal J} =0$, evaluated at complex locations in time and space, lead to singularities in Eulerian coordinates before the pre-shock?

To address this question, we show in the lower panel of Fig.\,\ref{fig:Lag1} the evolution of $\pm{\rm Im}(\mx(\mathfrak{a}_\pm, \mtau))$ as a function of~$|\mtau|$ for some selections of fixed phases $\Theta$, where $\mtau = |\mtau| \exp(\ii \Theta)$ (notice that we reserve the letter $\theta$ for the phase of the singularity).
It is crucial to observe  that the imaginary part of~$\mx$ vanishes at a $\Theta$-dependent value of $|\mtau|$.
Specifically,
for $\Theta=0$, the imaginary part of~$\mx$ vanishes at $|\mtau|=1$ which coincides precisely with the time of pre-shock.
By contrast, for $\Theta=\pi/2$, the imaginary part of~$\mx$ vanishes already at $|\mtau| \simeq 0.66274$, which agrees with the  estimate of $R_{\rm inf}$ from section~\ref{sec:asy-euler} to a precision of order $10^{-5}$.

\begin{figure}[t]
 {\centering
   \includegraphics[width=\columnwidth]{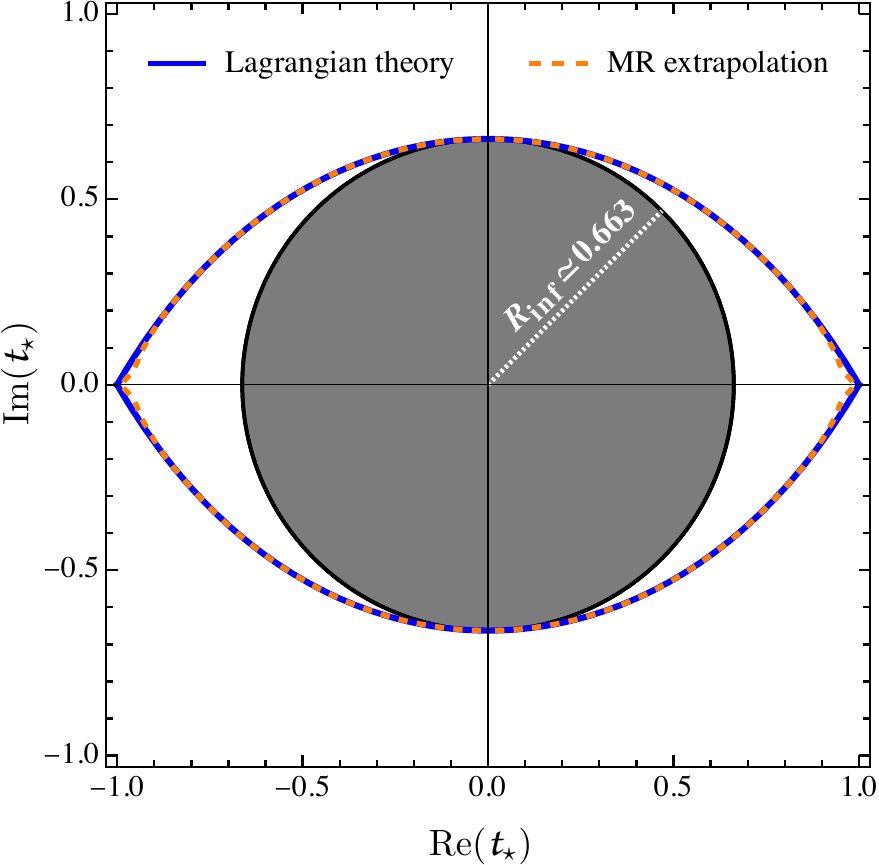}
  }
   \caption{Eye of the tyger, showing the locations of complex-time singularities $\mtstar= R\ws \exp(\ii \theta)$, based on the Lagrangian theory (blue solid line) and on the MR extrapolation technique (equations~\ref{eqs:MRstandardextra}; orange dashed line). The small discrepancy in the predictions stems from the aforementioned problem of the MR extrapolation technique becoming slightly inaccurate when $\theta \approx 0, \pm \pi$. }   \label{fig:Lag2}
\end{figure}

From these considerations it becomes clear that it is indeed the vanishing of the Jacobian in Lagrangian space, evaluated at complexified  locations, that is responsible for the birth of the  early-time tygers in the Eulerian space. More precisely, to search for the complex-times $\mtau_\star$ that lead to singularities  at the real-valued Eulerian position~$\mx$, we impose the vanishing of the imaginary part of $\mathpzc{x}(\text{\small $\mathfrak{a}\!=\!\mathfrak{a}_\pm$}, \text{\small $\mtau\!=\!\mtstar$})$. Specifically, using Eq.\,\eqref{eq:complexX}, we demand
\begin{align} \label{eq:constraint}
 \boxed{ \mtau =\mtstar  : \quad     {\rm Im}\! \left[ \arccos\left(\frac 1 \mtstar \right) -  \mtstar \sqrt{1-\frac 1 { \!\ws\mathpzc{t}_{\hspace{0.02cm}\star}^2\ts}} \,\ts \right] = 0 } 
\end{align}
(For multi-mode initial data, this condition has to be slightly generalized; see section~\ref{sec:multi}).
On a technical level, we vary parametrically $\Theta =\theta$ and determine the corresponding $|\mtstar|\! =\!R$ that satisfies the condition~\eqref{eq:constraint}, leading to $\mtstar\! =\! R \exp(\ii \theta)$ as a function of $\theta$. 
Then, by drawing parametrically $R(\theta)$ and $\theta$  against ${\rm Re}[\mx(\text{\fontsize{8}{8}\selectfont $\mathfrak{a}_\pm(\mtstar\text{\fontsize{7}{7}\selectfont $(\theta)$}),\mtstar(\theta)$})]$, one obtains respectively the solutions as shown  in the top and bottom panels of Fig.\,\ref{fig:Rthetanu} (green solid lines).
In Fig.\,\ref{fig:Lag2} we show $\mtstar$ as predicted from~\eqref{eq:constraint}, and compare it against the results from the MR extrapolation technique (section~\ref{sec:asy-euler}). 
Overall, the agreement between theory and extrapolation is excellent, except for complex values of $\mtstar$ that are close to the real time axis. The reason for this discrepancy is the aforementioned problem of the extrapolation becoming slightly inaccurate when~$\theta \to 0, \pm \pi$.

As an application of the above, we show  how our theory can be used to determine the well-known $3/2$ exponent of the pre-shock singularity \cite{zbMATH03915880,SULEM1983138,2011PhRvE..84a6301R}.
The pre-shock occurs at the real time $t_\star =1$ and real location $x=0$ with $\theta=0$, in which case Eq.\,\eqref{eq:complexX} reduces to $\mathpzc{x}= \pm [\arccos(1/t) - \sqrt{t^2-1}]$. Setting in this equation $t = t_\star - \delta t$ and Taylor expanding around the discrepancy $\delta t>0$, one finds 
\begin{align}
  \mathpzc{x}_{\,\star} &= \pm \ii \frac{2\sqrt{2}}{3}  \delta t^{3/2}  \pm \ii  \frac{3}{5 \sqrt{2}} \delta t^{5/2}    + O(\ii \delta t^{7/2})\,,
\end{align}
i.e., singular behaviour that is perfectly aligned along the imaginary axis.
To our knowledge, the sub-leading asymptotic behaviour with exponent $5/2$ has not yet been reported in the literature. Of course, using our theory, the asymptotic behaviour could be analyzed to arbitrarily high level.

Concluding this section, we have seen that temporal singularities in the inviscid Burgers' equation can be detected by essentially exploiting its exact Lagrangian-coordinates solution until pre-shock, i.e., the Lagrangian map. It may come as a surprise how singularities can arise within this (seemingly singularity-free) description. However, once the map is evaluated at the Lagrangian roots $\mathfrak{a}_\pm$ associated with the pre-shock, square roots are introduced (cf.\ Eq.\,\ref{eq:complexX}). As a consequence, derivatives of the map, evaluated at $\mathfrak{a}=\mathfrak{a}_\pm$, are singular for~$\mtau = \mtstar$.

Finally, Platzman's Eulerian solution for the velocity (Eq.\,\ref{eq:uPlatz}) could be the starting point of a similar singularity analysis as outlined above. Indeed, based on numerical tests, we have obtained evidence that Platzman’s solution displays non-convergent behaviour when evaluated at sufficiently large complex times. For example, non-convergence is observed if one evaluates the second time derivative of Eq.\,\eqref{eq:uPlatz} at $x=\pi/2$ and complex time $\mtau =  \mtstar(\text{\small $x=\pi/2$}) \simeq 0.663 \ii$ (cf.\ Fig.\,\ref{fig:Rthetanu}), which is precisely in line with the above analysis.
This singular behaviour can also be understood by the explicit structure of Platzman's solution. For this observe that the r.h.s.\ of Eq.\,\eqref{eq:uPlatz} is comprised of sums of products of Bessel and sine functions. While both Bessel  (for integer indices) and sine functions are entire functions in their arguments,  an infinite sum of products of piecewise entire functions does generally  have singularities in the complex domain (see e.g.\ \cite{ablowitz_fokas_2003}).

\section{Strategies for halting Tygers} \label{sec:arrestETTs}

From the above analysis it is clear that convergence, and thus, the range of time-analyticity, of $u = \sum_{n=0}^\infty u_n \ws t^n$ is severely hampered by the emergence of complex-time singularities, which is accompanied by the birth of early-time tygers.
The natural question is then, how the regime of time-analyticity could be extended along the real time axis---which is the physically relevant branch.

One obvious way is to exploit exact analytical results, such as the one in Lagrangian coordinates (Eq.\,\ref{eq:ZAmap}), or the one of Platzman for the case of single-sine-wave initial conditions (Eq.\,\ref{eq:uPlatz}) for $0\leq t < t_\star$. However, exact results for inviscid fluid equations, in particular also in higher spatial dimensions, are challenging to find.

Another way to extend the range of analyticity is to apply an analytic continuation technique (\`a la Weierstrass) within the time-Taylor series approach, such that a sequence of times $0 < t_1 < t_2 < \ldots$ can be constructed with $|t_{n+1}-t_n| < R(t_n)$, where $R(t_n)$ is the radius of convergence around the expansion point $t_n$ (see e.g.\ \cite{2016JCoPh.306..320P,2015MNRAS.452.1421R} for related approaches).  In words, an extended range of analyticity can be constructed by a multi-time-stepping procedure, where each time-step is strictly smaller than the current radius of convergence. Such an analytic continuation is amenable until the time of pre-shock, which is a real singularity, and thus forbids any further continuation beyond the pre-shock.

Such an analytic-continuation technique is however fairly elaborate, as at each time step one would need to find (at least roughly) the radius of convergence around the current expansion point. 
Furthermore, as one approaches the pre-shock singularity located on the real time-axis, the current radius of convergence becomes naturally very small, thereby allowing only incremental time steps. Thus,  analytic-continuation techniques can become very inefficient if singularities are close to the real axis, and hence we do not follow such an approach in the present paper.

Instead, here we report two  methods that allow us to halt efficiently the tygers in a single-time step. One of the methods is tyger purging, a variant of the numerical method developed by Ref.~\cite{2020PhRvR...2c3202M}, which essentially removes ultra-violet features from the velocity (section~\ref{sec:purge}). The other is a novel technique inspired by Duhamel's principle, and instead attempts to tame tygers by performing a (partial) ultra-violet completion (section~\ref{sec:UVcomplete}).

\subsection{Tyger purging}\label{sec:purge}

\begin{figure*}[t]
 {\centering
   \includegraphics[width=\textwidth]{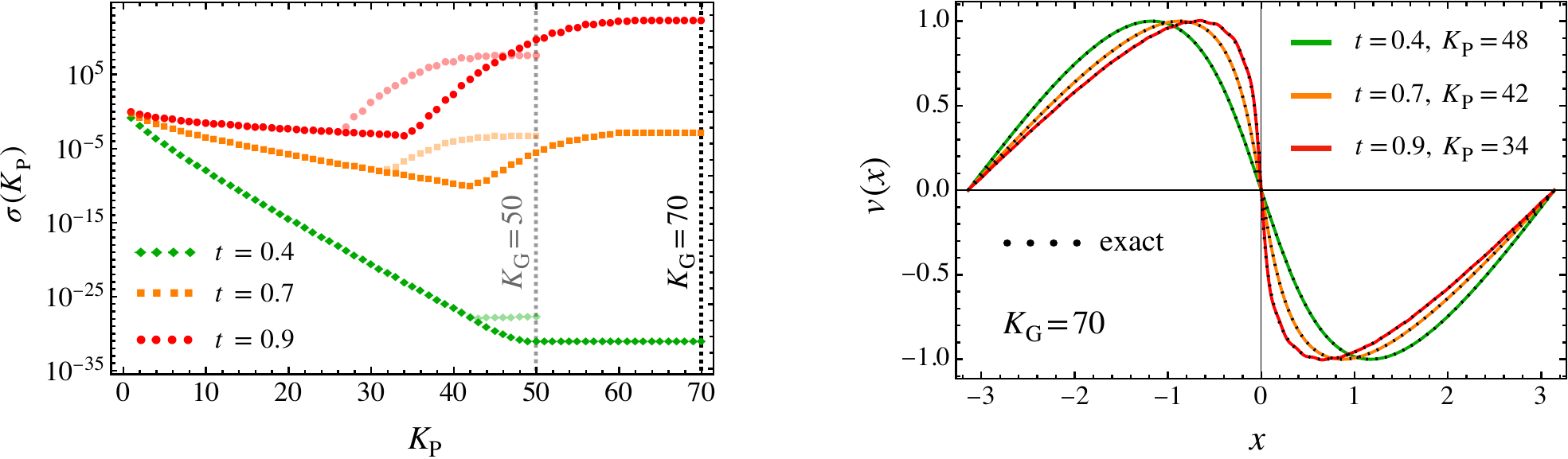}
  }
   \caption{Analysis of tyger purging for the Taylor-series approach with single-mode initial conditions. {\it Left panel:} Integrated error $\sigma$ (equation~\ref{eq:sigmaerror})  as a function of $\KP$, where the modes $\KP < |k| \leq \KG$ are removed from the velocity. Shown is $\sigma$ for $t=0.4,0.7,0.9$ (green, orange and red symbols respectively) for the truncation $\KG = 70 = N+1$ (faint lines: $\KG = 50$). For sufficiently small $\KP$ and times $t < t_\star=1$, the integrated error is exponentially decaying. For very early times, the maximal precision levels off at a $\KG$-dependent threshold for $\KP$, while for later times after $t \gtrsim R_{\rm inf} = 0.663$, we observe an increase in $\sigma$ for large $\KP$, indicating a significant loss in precision due to the birth of tygers. {\it Right panel:} The velocity with optimal purging strategy where $\sigma$ is minimal. }   \label{fig:purging}
\end{figure*}

For the single-mode initial conditions~\eqref{eq:ICsinglemode}, we have seen that the $N$th-order Taylor coefficient of the velocity is band-limited with the highest wave number being $k= \pm(N+1)$.
Thus, if we represent the truncated Taylor series in Fourier space, then the corresponding Fourier representation is naturally Galerkin truncated at wave number $|k| = \KG = N+1$, i.e., 
\be
  v :=  \PNof{N} u  = \sum_{k=0,\pm 1, \pm 2, \ldots, \pm (N+1)} \hat u_k \ws  {\rm e}^{\ii k x} 
\ee
(The effective Galerkin truncation depends on the form of the initial data; see e.g.\ section~\ref{sec:multi} for the case of two-mode initial data).
Such a Galerkin-truncated Fourier representation is also employed in fully numerical
avenues of inviscid fluid equations (e.g.\ \cite{doi:10.1137/0726003,2005PhRvL..95z4502C,PhysRevE.87.033017,2018PhRvF...3a4603C}). However, in the  Taylor-series approach with trigonometric initial conditions, all Fourier coefficients $\hat u_k$ can be determined explicitly, i.e., without resorting to a numerical mesh that would come with various approximations.

Thus, the Taylor-series approach provides us with a clean setup for applying the tyger purging method of Ref.~\cite{2020PhRvR...2c3202M}, albeit certain modifications are necessary given the different nature of the approaches.
First of all, \cite{2020PhRvR...2c3202M} suggests to remove Fourier modes in a narrow band around the Galerkin truncation $\KG$ of the numerical mesh, specifically the range $\KP < |k| \leq \KG$,  where the optimal choice of $\KP \simeq \KG - \KG^{0.8}$ has been confirmed by various numerical experiments in the range $\KG = 500 - 10000$. 
As mentioned above, we do not need to sample our solutions on a numerical mesh, but the truncation order in the Taylor series acts as an effective Galerkin truncation in our approach.
At the same time, the effective Galerkin truncation for the Taylor-series approach is typically significantly smaller compared to the one used for numerical simulations.
Therefore, we expect that the optimal choice for $\KP$ should be different in the present approach as reported in Ref.\,\cite{2020PhRvR...2c3202M}.
Second, \cite{2020PhRvR...2c3202M} applied also a purging strategy to the time-step sizing for the temporal integration, which is however in the present approach not needed as we only consider single time steps, i.e., we evolve directly from initial time to final time. Nonetheless, as we will see, the actual value of the final time influences the choice of how many modes should be removed. Loosely speaking, the later the final time, the stronger the amplitude of the tygers, and the more modes that need to be discarded.

To proceed, we define the purging operator $\PKP$ which removes the Fourier modes in the band $\KP < |k| \leq \KG$ from the velocity, i.e.,
\be
 \PKP v(x)  = \sum_{|k| \leq K_{\!\ws\tiny \text{P}}} \hat u_k \ws {\rm e}^{\ii k x} \,.
\ee
To analyze the impact of this operator, we further define the integrated error with respect to the exact solution, i.e., 
\be \label{eq:sigmaerror}
  \sigma(t, \KP) := \int_{-\pi}^{+\pi} \left[ \PKP v(x(a,t),t) - u_0(a) \right]^2 \dd a \,.
\ee
In the left panel of Fig.\,\ref{fig:purging}, we show the integrated error as a function of the low-pass threshold $\KP$, for $t=0.4, 0.7$ and~$0.9$ (shown respectively in green, orange and red) while setting $\KG =70 = N+1$ [faint lines: $\KG=50$]. For all considered times, the integrated error decays roughly exponentially for low values of $\KP$, albeit with a significantly flattening slope at subsequent later times. This is a generic behaviour that is  expected for a Taylor series that is evaluated in the vicinity of its radius of convergence.
At very early times ($t=0.4$), the integrated error levels off at a $\KG$-dependent precision for $\KP \geq 48$ [faint lines: $\KP \geq 42$]
but, importantly, the integrated error remains constant. 
Thus, higher-order modes could be added without harming the accuracy of the results.

However, at times $t \gtrsim R_{\rm inf} \simeq 0.663$, we observe in the left panel of Fig.\,\ref{fig:purging} a sudden increase in amplitude of $\sigma$ for large $\KP$,  which is, as we claim, driven by the birth of tygers.
From these considerations it becomes evident  that an optimal purging strategy is achieved   for the maximal value of~$\KP$ for which the integrated error is minimal.
Specifically, for the shown times $t=0.7$  and $t=0.9$, the maximal precision is achieved for $\KP =42$ and $\KP =34$, respectively [faint lines: $\KP =31$ and $\KP =26$]. More generally,  we find the fitting function $\KP \simeq \KG - 0.61 (\KG -4.9)t$  to be accurate in the tested range $\KG = 30- 70$.

In the right panel of Fig.\,\ref{fig:purging}, we show the velocity with the outlined purging strategy. The agreement with the exact solution (black dotted lines) is in general very good, especially considering that the times $t=0.7, 0.9$ are well beyond $R_{\rm inf} \simeq 0.663$. 
This indicates that the purging strategy removes the impact of complex-time singularities on the Taylor truncation of the velocity.
However, for very late times, there are signs of loss of precision \mbox{($t=0.9$,} red line). A more accurate  solution close to the pre-shock could be obtained by going to (significantly) higher Taylor orders in the velocity, followed by an appropriate updated purging strategy. We leave such avenues for future works.

\subsection{Iterative UV completion} \label{sec:UVcomplete}

We have seen that removing Fourier modes below the Galerkin (Taylor) truncation does tame early-time tygers.  Here we raise the question whether something similar could be achieved by {\it adding} Fourier modes beyond the original Galerkin truncation.

To assess such a possibility,  we reconsider Burgers' equation, which can be written in conservative form as
\be \label{eq:Burgerscons}
  \partial_t u =  - \frac 1 2 \partial_x  u^2 \,.
\ee
Of course, in the smooth case, the formal solution of~\eqref{eq:Burgerscons} can be obtained by integration from $0$ to $t$:
\be \label{eq:BurgersInt}
  u = u_0 - \frac 1 2 \partial_x \int_0^t u^2(\tau) \,\dd \tau \,.
\ee
Here, $u_0$ is the initial velocity, and from now on, we occasionally suppress the spatial dependence when there is no source of confusion.

Let us approximate the quadratic term in Eq.\,\eqref{eq:BurgersInt} by replacing $u^2 = (\PN u)^2$ where $\PN u$ is,  as before, the Taylor-series representation of the velocity at truncation order $N$. The resulting approximation for the velocity is called $\mathpzc{v}_{\{1\}}$ and governed by
\begin{align}
  \mv{1} &= \mvini  - \frac 1 2 \partial_x \int_0^t [\PN u(\tau)]^2 \, \dd \tau \,, 
  \qquad \mvini  = u_0 \,.
\intertext{Here, and similarly for higher iterations, we drop the implicit dependence of $\mv{1}$ on $N$ for conciseness.
Now, we propose a  {\it bootstrapping method} to this equation such that a (possibly) refined approximation of $\mv{1}$ is obtained by replacing the quadratic term in the integrand by $\mv{1}^2$. We call the resulting approximation $\mv{2}$, and the governing equation is}
  \mv{2} &= \mvini  - \frac 1 2 \partial_x \int_0^t \mv{1}^2(\tau)\, \dd \tau \,.
\intertext{Of course, such a bootstrapping can be continued iteratively, so in general we can write for the $i$th bootstrapped solution for the velocity ($i>2$)}
  \mv{i} &= \mvini  - \frac 1 2 \partial_x \int_0^t \mv{i-1}^2(\tau)\, \dd \tau \,. 
\end{align}
The outlined method has at least two intriguing features: 
First, the bootstrapping only adds new modes and new Taylor coefficients beyond the original truncation-order $N$ that were not already present in the input.
Second, the method is very efficient in populating Fourier modes in the UV regime. Specifically, for the given single-mode initial data, the Taylor-series input $\PN u$ has no wave numbers beyond $k= \pm(N+1)$, while $\mv{i}$ has the highest wave number at $k= \pm {\cal N}_i$  with ${\cal N}_i = 2^i  ( N+1 ) -1$.

\begin{figure}[t]
 {\centering
   \includegraphics[width=0.93\columnwidth]{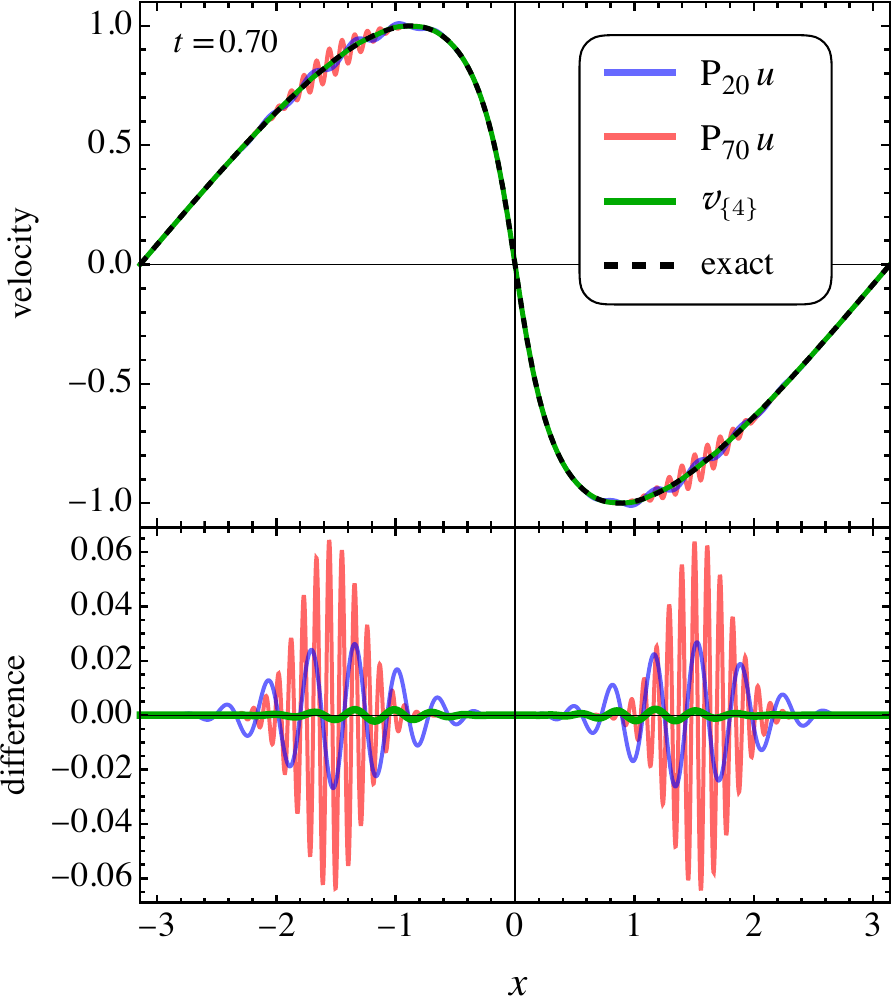}
  }
   \caption{{\it Top panel:} The bootstrapped velocity $\mv{4}$ (green line) compared against the exact solution (black dashed line), as well as against 
    the truncations $\PNof{20}u$ (blue line) and $\PNof{70}u$ (pink line). We remark that none of the shown predictions have been purged. {\it Bottom panel:} The difference ${\cal U} - u$, where ${\cal U}$ are the velocity predictions from the top panel with the same color code, and $u$ is the velocity of the exact solution. Evidently, the tyger amplitude is strongly suppressed for the bootstrapped velocity in comparison to its input $\PNof{20}u$.
   }   \label{fig:UVcomplete}
\end{figure}

In the top panel of Fig.\,\ref{fig:UVcomplete}, we show $\mv{4}$ (green line) which we have generated using as the input the  truncated Taylor-series $\PNof{20}u$ (blue line) with single-mode initial condition~\eqref{eq:ICsinglemode}.
Based on the above formula, this bootstrapped solution contains Fourier modes up to 
$k =\pm 335$.  We compare the bootstrapped solution against the exact solution (black dashed line), as well as against $\PNof{70}u$ (red line) which 
contains ``only'' modes up to $k=\pm 71$, and it is the highest truncation order considered in the present paper. While $\PNof{20}u$ and in particular $\PNof{70}u$ exemplify the birth of tygers, the bootstrapped solution appears almost tyger-free. 
To be specific, we list in Tab.\,\ref{tab:tyger} the maximal tyger amplitudes 
\be
  {\cal T}({\cal U}) := {\rm max}_x \left|\, {\cal U}(x,t) - u(x,t) \right|
\ee 
for various truncated {\it and unpurged} solutions~${\cal U}$, where $u$ is the velocity of the exact solution. It is seen that, depending on the considered time, the tyger amplitudes shrink roughly by a factor of $1.3-1.9$ for each iteration within the bootstrapping method.
Thus, the bootstrapping adds successively higher modes into the UV regime that appears to partially cure the truncated Taylor-series associated to early-time tygers.
However, even with the highest considered bootstrapped solution, $\mv{4}$, the loss of accuracy becomes unsatisfactory around times~$t \gtrsim 0.75$. This could be possibly rectified by employing higher iterations in the bootstrapping; see also the final paragraph at the end of this section.

\newlength\q
\setlength\q{\dimexpr 0.071\textwidth -2\tabcolsep}

{\renewcommand{\arraystretch}{1.5}

\begin{table}
\caption{Maximal tyger amplitude $\cal T$ at times  $t  > R_{\rm inf}$. The bootstrapped solutions $\mv{1,2,3,4}$ were generated using $\PNof{20}u$ as the input.}
\centering
\begin{tabular}{l|p{\q}p{\q}p{\q}p{\q}p{\q}p{\q}}    
\hline
 Time & $\mT{\PNof{70}u}$ & $\mT{\PNof{20}u}$ & $\mT{\!\mv{1}\!}$ & $\mT{\!\mv{2}\!}$ & $\mT{\!\mv{3}\!}$ & $\mT{\!\mv{4}\!}$ \\ \hline
 0.70 & 0.0645 & 0.0268  & 0.0141 & 0.0076 & 0.0041 & 0.0022 \\ 
 0.75 & 7.7795 & 0.1030  & 0.0573 & 0.0333 & 0.0188 & 0.0104 \\ 
 0.80 & 680.11 & 0.3617  & 0.2155 & 0.1395 & 0.0814 & 0.0447 \\ 
 0.85 & 46175. & 1.1747  & 0.8391 & 0.6551 & 0.3895 & 0.1934 \\ 
\hline 
\end{tabular}
\label{tab:tyger}
\end{table}

\begin{table}
\caption{Same as Tab.\,\ref{tab:tyger} but shown is the quantity $\delta \!\hspace{0.02cm} E$ which parameterizes the violation of the energy conservation.}
\centering
\begin{tabular}{l|llllll} 
\hline
Time& $\mE{\PNof{70}u}$ & $\mE{\PNof{20}u}$ & $\mE{\!\mv{1}\!}$ & $\mE{\!\mv{2}\!}$ & $\mE{\!\mv{3}\!}$ & $\mE{\!\mv{4}\!}$ \\ \hline
  0.70 & 6.14e$\mminus$4 & 2.03e$\mminus$4  & 5.52e$\mminus$5 & 1.49e$\mminus$5 & 4.06e$\mminus$6 &  1.10e$\mminus$6 \\ 
 0.75 &  8.93e+0  & 3.00e$\mminus$3  & 9.09e$\mminus$4 & 2.77e$\mminus$4 & 8.56e$\mminus$5 &  2.65e$\mminus$5 \\ 
 0.80 &  6.97e+4  & 3.70e$\mminus$2  & 1.25e$\mminus$2 & 4.26e$\mminus$3 & 1.47e$\mminus$3 &  5.12e$\mminus$4 \\ 
 0.85 & 3.14e+8 & 3.90e$\mminus$1 &  1.56e$\mminus$1   &6.09e$\mminus$2 & 2.15e$\mminus$2 & 8.24e$\mminus$3 \\ 
\hline 
\end{tabular}
\label{tab:energy}
\end{table}

Very similar statements can be made about the energy, which should be conserved before the time of pre-shock.
To assess this crucial point, we define the error on the energy conservation
\be \label{eq:energy}
  \delta \!\hspace{0.02cm} E({\cal U}) := \frac{2}{\pi} \int_{-\pi}^{+\pi} \frac{{\cal U}^2(x,t)}{2} \dd x  - 1 \,,
\ee
which is exactly zero if the energy for the velocity ${\cal U}$ is conserved.
Clearly, as long as the time-Taylor series for the velocity converges, $\mE{\PNof{N}u}$ should tend to zero for increasingly large truncation-orders~$N$. 
However, for $|\mtau| \geq R_{\rm inf}$, i.e.,  when convergence is lost, the situation is drastically different: the Taylor series diverges, which is accompanied by the violation of energy conservation.
This is demonstrated in Tab.\,\ref{tab:energy} for the two Taylor truncations $\PNof{20}u$ and $\PNof{70}u$. In the same table,  we also report the errors $\mE{\!\mv{i}\!}$ for the bootstrapped solutions: In stark contrast, it is seen that the errors $\mE{\!\mv{i}\!}$  are decreasing for increasing iterations $i=1,2,3,4$, at all considered times. In particular, $\mE{\!\mv{4}\!}$ is roughly two orders of magnitudes smaller than the error $\delta \!\hspace{0.02cm} E$ of its input $\PNof{20}u$.
Thus, the bootstrapping partially restores the conservation of energy; we will return to this matter in our conclusions.
However, from Tab.\,\ref{tab:energy} it is also seen that the iterative bootstrapping becomes increasingly inefficient for times closer to the pre-shock.

We speculate that there might be a non-perturbative resummation of the outlined bootstrapping method, very much in the sense as it is known in quantum electrodynamics, where the Neumann series for the time evolution operator can be recast into the Dyson series (e.g.\ \cite{sakurai_napolitano_2017}). Such a resummation of the bootstrapping method, if available, could then be viewed as a ``full'' UV completion.

\section{Analysis for multi-mode initial data}\label{sec:multi}

\subsection{Phenomenology, purging and convergence}

Here we provide phenomenological and theoretical results for the two-mode initial data 
\be \label{eq:twomodeIC}
   u_0 =   - \sin x - 4 \cos(2x) \,,
\ee
for which the first pre-shock occurs at $t_\star \simeq 0.1147$ at $x = \xstar \simeq -0.7043$. 
To our knowledge, no exact Eulerian-coordinates solutions are known in the present case (cf.\ Eq.\,\ref{eq:uPlatz}, valid for single-sine-wave initial data). 
Using instead the recursive relation~\eqref{eq:rec}, one can easily generate the Taylor series coefficients of $u = \sum_{n=0}^\infty u_n\ws t^n$. 
In the present paper, we have determined 40 Taylor coefficients for the two-mode initial data; the first coefficients and the $N$th coefficient read 
\begin{subequations}
\begin{align}
  u_1 &=    \cos(x) [2 - \sin(x) ] - 6 \cos(3 x) + 16 \sin(4 x)  \,, \\
  u_2 &=   \frac{33}{8} \sin(x) + 36 \cos(2 x)  -  \frac{147}{8} \sin(3 x) - 8 \cos(4 x)   \nonumber \\
     &\qquad  + 50 \sin(5 x)  + 96 \cos(6 x)\,, \\
   &\,\,\, \rotatebox{90}{$\cdots$}  \nonumber \\
  \uN &= \cdots + \calcN {\rm tri}(2[N+1]x) \,, \label{eq:un-twomode}
\end{align}
\end{subequations}
where $\calcN$ is a constant, and ${\rm tri}$ is a sine [cosine] when $N$ is odd [even]. Note that, in comparison to the single-mode case which contains modes  $|k| \leq  N+1$ at truncation order~$N$, for the present multi-mode initial data we have $|k| \leq 2(N+1)$. As a consequence, the purging strategy has to be trivially updated.

\begin{figure}[!t]
 {\centering
   \includegraphics[width=0.99\columnwidth]{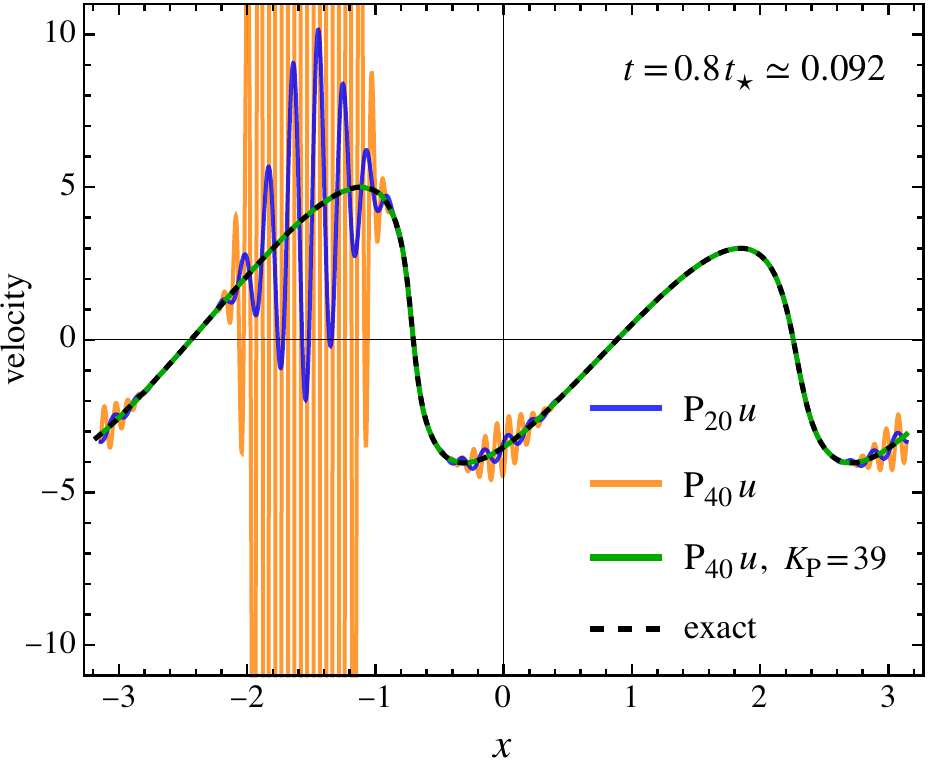}
  }
   \caption{The velocity for the two-mode initial data~\eqref{eq:twomodeIC} at $t = 0.8 t_\star$, where the time of first pre-shock is $t_\star \simeq 0.1147$. Shown are the truncated solutions $N=20,40$ (blue and orange lines), which exemplify the birth of one dominant and two subdominant tygers centred respectively around $x \simeq -\pi/2,\, 0.0527$ and $3.0889$ (the precise positions are obtained from Fig.\,\ref{fig:R-theta-twomode}). 
Also shown is the truncated solution with $N=40$, after applying the optimal purging strategy with $\KP =39$ (green line), which agrees well with the exact solution (black dashed line). There is yet another tyger at $x=+\pi/2$ that however only appears at times $t\simeq 0.1$; see text.
}   \label{fig:phasespace-twomode}
\end{figure}

Figure~\ref{fig:phasespace-twomode} shows the phase-space for the two-mode initial data~\eqref{eq:twomodeIC} at $t = 0.8 t_\star$. It is seen that tygers are born at multiple locations, with the strongest tyger centered around $x=-\pi/2$. Interestingly and in contrast to the single-mode case (cf.\ Fig.\,\ref{fig:phasespace}), there appears to be no tyger (yet) at the location $+\pi/2$, at least  not for the considered time $t = 0.8 t_\star$; we will further comment on this shortly. Instead, two tygers with smaller amplitude in the neighborhood of $x\simeq 0.0527$ and $3.0889$ are born.

\begin{figure}[t]
 {\centering
   \includegraphics[width=0.92\columnwidth]{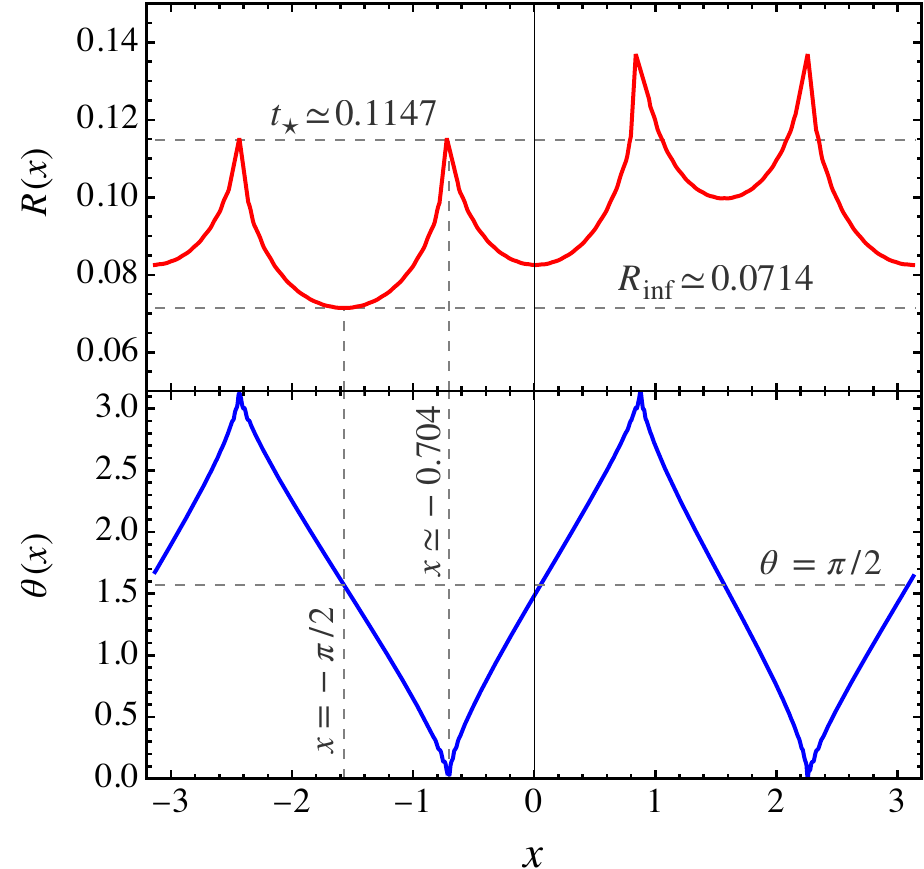}
  }
   \caption{Locations of singularities $\mtstar = R \exp(\ii \theta)$ for the case of the two-mode initial data~\eqref{eq:twomodeIC}, obtained from the MR extrapolation technique based on Eqs.\,\eqref{eq:Bn}--\eqref{eq:costhetan}.  Four distinct locations are found where $\theta =\pi/2$, which are linked to the birth of early-time tygers. The most dominant tyger is the one at $x=-\pi/2$ and will appear around $t \simeq 0.0714$, which coincides with $R_{\rm inf}$. Subdominant tygers appear at later times $t \simeq 0.0825$ around the locations $x\simeq 0.0527$ and $3.0889$, while yet another tyger will be born  at $t \simeq 0.100$ located at $x=\pi/2$  (not yet visible in~Fig.\,\ref{fig:phasespace-twomode}). The horizontal lines marking $t_\star$ and $R_{\rm inf}$ are theoretical values obtained through the Lagrangian theory (section~\ref{sec:Lag-twomode}).
}   \label{fig:R-theta-twomode}
\end{figure}

To address when and where tygers are born, let us analyze the convergence of the Taylor series by applying the identical Mercer--Roberts methodology as outlined in section~\ref{sec:asy-euler} (equations~\ref{eq:Bn}--\ref{eq:costhetan}).
In Fig.\,\ref{fig:R-theta-twomode} we show the resulting estimates for the radius of convergence (top panel), as well as the corresponding phase of the singularity (bottom panel).
Using the nonlinear extension of the MR extrapolation technique (see paragraph after Eq.\,\ref{eq:costhetan}), we find $R = 0.115$ at pre-shock location, which agrees with $t_\star$ to a precision of $0.26\%$.

For the location $x\!=\!-\pi/2$ for which the strongest tyger is expected, the MR extrapolation yields $\theta(\text{\small $x\!=\!-\pi/2$}) \simeq 1.57078$ which agrees with $\pi/2$ to a precision of $10^{-5}$. 
Thus, as expected, at $x\!=\!-\pi/2$ there is a purely imaginary singularity linked to the strongest tyger as shown in Fig.\,\ref{fig:phasespace-twomode}. Regarding the local radius of convergence at  $x\!=\!-\pi/2$, the MR technique reveals $R \simeq 0.0714$ which agrees with the theoretical value of $R_{\rm inf}$ to a precision of $0.075\%$ (see section~\ref{sec:Lag-twomode} for the theoretical results).
Similarly, there are purely imaginary singularities at the spatial locations $x\!\simeq\!0.0527,3.0889$ which are associated with the aforementioned subdominant tygers. Based on the upper panel in Fig.\,\ref{fig:R-theta-twomode}, we can even deduce why these tygers have a smaller amplitude in comparison to the dominant tyger: the corresponding radius of convergence at the locations $x\!\simeq\!0.0527$ and~$3.0889$ is slightly larger, namely $R\simeq 0.0825$, as compared to $R_{\rm inf} \simeq 0.0714$ at $x\!=\!-\pi/2$. Thus, the dominant tyger started growing already at $t \simeq 0.0714$ and hence had more time to grow, as compared to  the subdominant tygers that  appear around $t\simeq 0.0825$.

Finally, by similar arguments as outlined above, one would expect also a tyger appearing at the location $x=+\pi/2$ which is not yet visible in Fig.\,\ref{fig:phasespace-twomode}. 
Indeed, at $x=\pi/2$, we find $R \simeq 0.0100$ and $\theta \simeq \pi/2$ to a precision of $10^{-6}$,
and we have explicitly verified that the corresponding tyger shows up at this location for times around $t \simeq 0.100$.

\subsection{Lagrangian singularity theory for multi-mode initial data}\label{sec:Lag-twomode}

Here we apply the Lagrangian singularity theory 
of section~\ref{sec:asy-lagrange}  
to the two-mode initial data~\eqref{eq:twomodeIC}; the generalization to the multi-mode case is straightforward and discussed at the end of the section.
 Employing the direct Lagrangian map \mbox{$a \mapsto x$,} one finds 
\begin{align}
  x(a,t) = a - t \left[ \sin a + 4 \cos(2a)\right] \,,
\end{align}
which implies the Jacobian determinant 
\be
  J(a,t) = 1 + t \left[ 8 \sin(2a) - \cos a  \right] \,. 
\ee
From these solutions it is elementary to determine the time of the first pre-shock, 
\begin{align}
 t_\star &= \sdfrac{1}{255} \sqrt{\sdfrac{176469- 683 \sqrt{2049}}{170}} \simeq 0.11475  \,,
\intertext{as well as the Lagrangian location  of the pre-shock,}
  a_\star &=  - \arctan\left( \sqrt{\sdfrac{513 - \sqrt{2049}}{510} } \right) \simeq -0.76378 \,.
\end{align}
These results can  also be used to determine the current location of the pre-shock, dubbed $\xstar := x(a_\star,t_\star)$; the corresponding solution is explicit but lengthy, therefore we only provide its numerical value, $\xstar \simeq  -0.7043$.

To analyze the temporal singularities that occur in Eulerian space at times already well before $t_\star$, we follow an almost identical strategy as outlined in section~\ref{sec:asy-lagrange}. 
Specifically, we again complexify the Lagrangian and Eulerian locations, dubbed respectively $\mathfrak{a}$ and $\mx$, and also allow the time variable to take complex values, i.e., $\mtau = |\mtau| \exp(\ii \Theta)$.

\begin{figure}[t]
 {\centering
   \includegraphics[width=0.92\columnwidth]{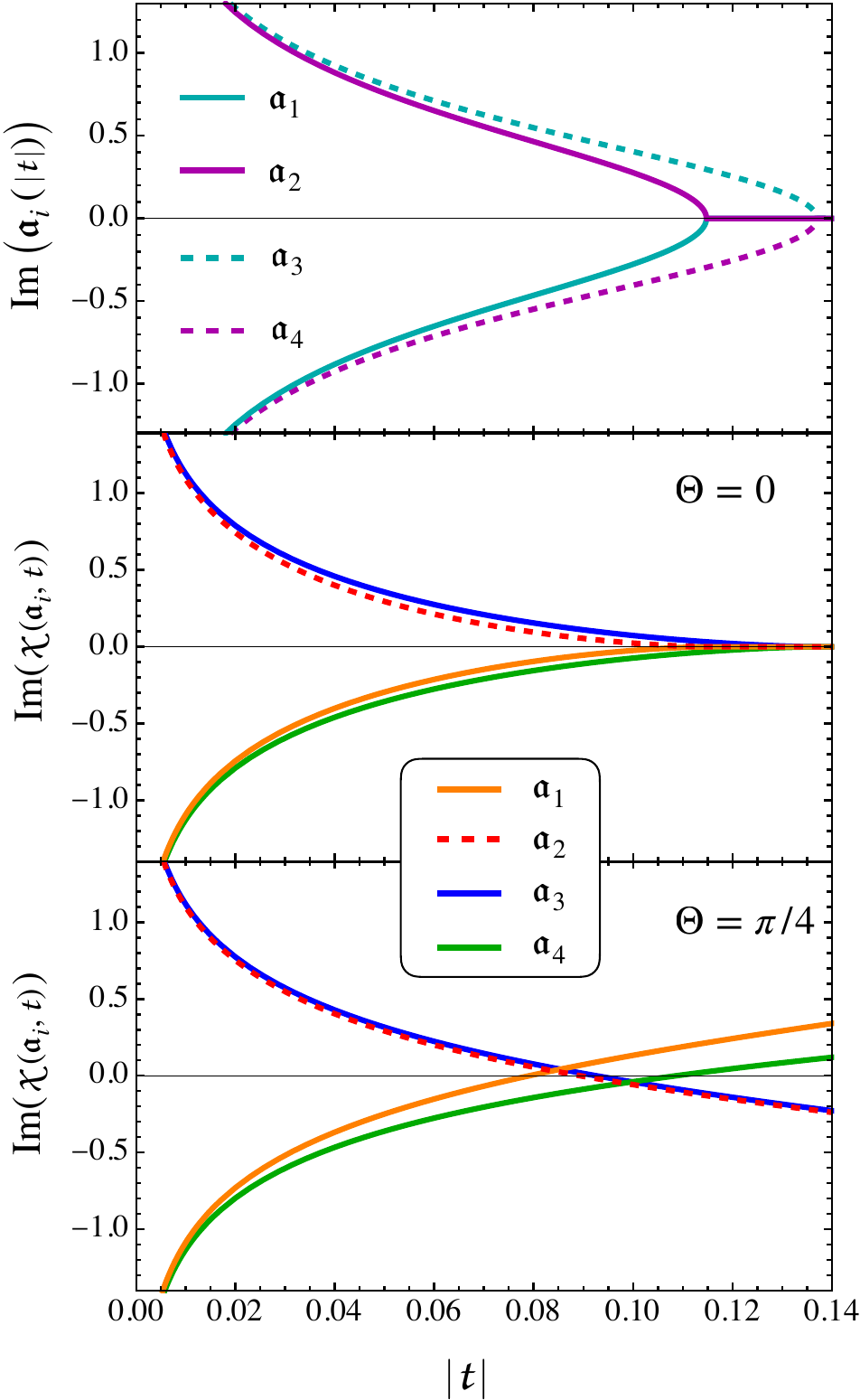}
  }
   \caption{Complex analysis of Lagrangian roots and map for the two-mode case.
{\it Top panel:} Temporal evolution of the imaginary part of the Lagrangian roots $\mathfrak{a}_i$ as a function of~$\mtau=|\mtau| \exp(\ii \Theta)$ with $\Theta=0$. 
  {\it Other panels:} Imaginary part of the complexified Lagrangian map, evaluated at the roots $\mathfrak{a}_i$, shown here for two choices of phases, namely $\Theta=0$ (central panel) and $\Theta= \pi/4$ (bottom panel). 
For multi-mode initial conditions, the physical relevant temporal singularity in Eulerian coordinates is determined by the smallest value of~$|\mtau|$ for which ${\rm Im}(\mx(\mathfrak{a}_i))$ vanishes for any of the roots~$\mathfrak{a}_i$. }  \label{fig:theory-twomode}
\end{figure}

We begin by searching for the Lagrangian roots $\mathfrak{a}_i$ for which the Jacobian vanishes at complex times. 
It is easily found that there are four such roots, $\mathfrak{a}_{1,2,3,4}$, which can be expressed in terms of explicit functions. The solutions  are however again very lengthy, therefore we show instead the temporal evolution of their imaginary parts in the top panel of Fig.\,\ref{fig:theory-twomode}. 
It is seen that the two roots $\mathfrak{a}_{1,2}$  have a vanishing imaginary part precisely at $|\mtau| = t_\star$ (shown as solid cyan and magenta lines). These two roots are thus associated with the first pre-shock. By contrast, the imaginary parts of the other two roots, $\mathfrak{a}_{3,4}$ (dashed lines colored in cyan and magenta respectively), vanish significantly later, indicating the appearance of a secondary pre-shock occurring at a time $|\mtau| \simeq 0.13697$ 
(but note that the present Lagrangian formulation ceases to be valid after the first pre-shock).

In Fig.\,\ref{fig:theory-twomode}, we also show the evolution of ${\rm Im}(\mx(\mathfrak{a}_i))$ as a function of $\mtau = |\mtau|\exp(\ii \Theta)$, specifically for two exemplary phases $\Theta\!=\!0,\,\pi/4$ (central and bottom panel, respectively).
 By a similar argument as given in section~\ref{sec:asy-lagrange},  it is precisely the vanishing of ${\rm Im}(\mx(\mathfrak{a}_i, \mtau))$  that leads to a singularity at the real-valued Eulerian position. In contrast to the single-mode case, however, we have now four distinct evolutions of ${\rm Im}(\mx(\mathfrak{a}_i, \mtau))$, depending on the selected roots $\mathfrak{a}_{1,2,3,4}$. Which of the root(s) should we select?

Physically, the most relevant singularity is the one that is closest to the origin in time (for a Taylor expansion around \mbox{$t\!=\!0$,} this is the singularity that sets the radius of convergence).
Thus, within a two-step process, we first define the critical times $\mtstar_{1,2,3,4}$ corresponding to the roots $\mathfrak{a}_{1,2,3,4}$, for which
\begin{subequations} \label{eqs:LagThMulti}
\be \label{eq:lag-cons-multi}
 \boxed{ \mtau =\mtstar_i  : \quad    {\rm Im} \left[ \mx(\text{\small $\mathfrak{a}\!=\!\mathfrak{a}_i$}, \text{\small $\mtau\!=\!\mtstar_{i}$} \right]  = 0 }
\ee
is satisfied. 
Then, as a second and final step, we select 
\be
  R  := \inf \left\{ |\mtstar_1| , |\mtstar_2| , |\mtstar_3|, |\mtstar_4|  \right\}  \,,
\ee 
\end{subequations}
which is the physically relevant radius of convergence~$R$ for fixed phase~$\Theta =\theta$.
This methodology is not only valid for the present two-mode case, but also generalizes obviously to the case of multi-mode initial conditions with an arbitrary number of Lagrangian roots.

\begin{figure}[t]
 {\centering
   \includegraphics[width=0.96\columnwidth]{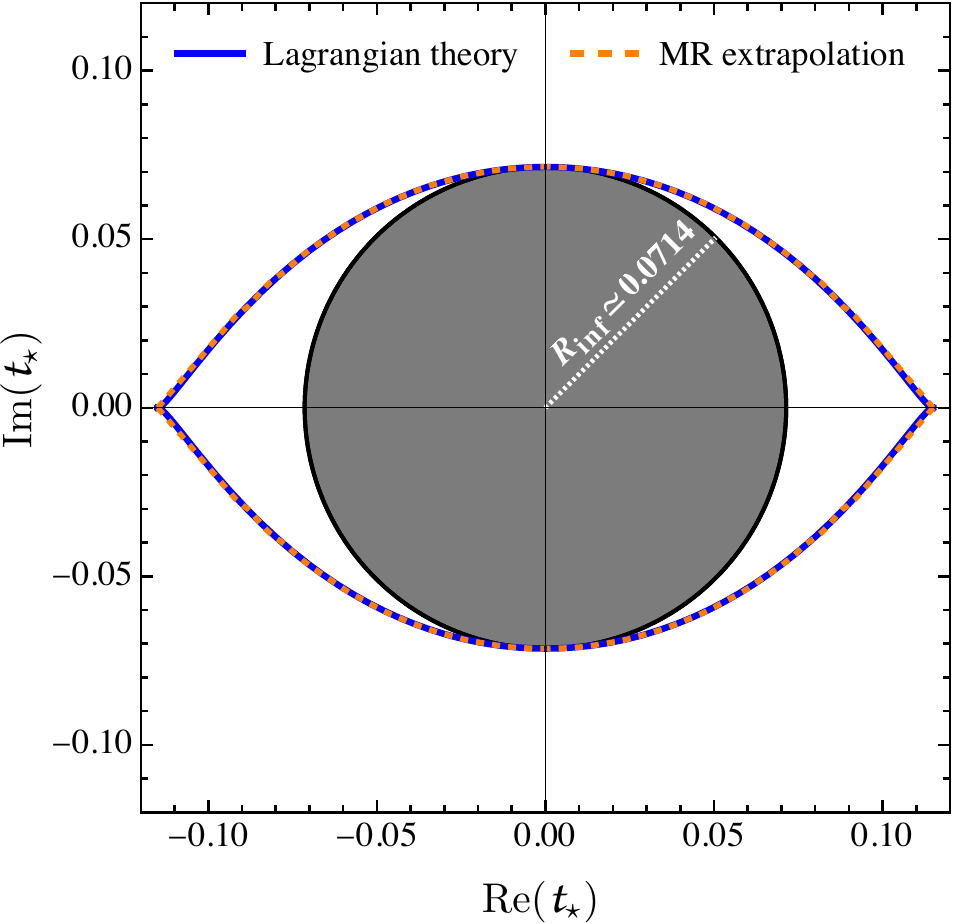}
  }
   \caption{Eye of the tyger for the two-mode case~\eqref{eq:twomodeIC}.  Shown are the {\it closest} complex-time singularities $\mtstar= R\ws \exp(\ii \theta)$, based on the Lagrangian theory (blue solid line; based on Eqs.\,\ref{eqs:LagThMulti}) and on the MR extrapolation technique (orange dashed line; Eqs.\,\ref{eqs:MRstandardextra}). 
}   \label{fig:eye-tyger-twomode}
\end{figure}

Finally, in Fig.\,\ref{fig:eye-tyger-twomode} we compare the theoretical results for~$\mtstar$ (solid blue line) against the predictions based on the MR extrapolation technique (orange dashed line; cf.\ Eqs.\,\ref{eqs:MRstandardextra}). 
In general, the results agree well, except when the singularities are located close to the real time axis: Similarly as in the single-mode case, this is the regime where the MR extrapolation technique becomes less accurate.

\section{Conclusions and perspectives}\label{sec:sum-out}

Until recently, tygers were observed in various numerical implementations of inviscid equations of hydrodynamics \cite{2011PhRvE..84a6301R,2017RSPSA,2018PhRvF...3a4603C,2017RSPSA,2020PhRvR...2c3202M,PhysRevE.87.033017}, 
and their mere existence may complicate the investigation of the blow-up problem in fluid dynamics (e.g.\ \cite{2008PhyD..237.1894G,doi:10.1137/140966411,HERTEL2022110758,ChenHou,2020arXiv201214182S}).
In such numerical setups, tygers appear initially as localized resonances  when complex-time singularities are near the real-time axis, which in the case of inviscid Burgers occurs at times close to the pre-shock, the first real singularity. 
Without any counter measures (such as purging \cite{2020PhRvR...2c3202M}), the amplitude and spatial width of tygers strongly increase in time, and eventually lead to a complete thermalization with an energy equipartition between all Fourier modes.

While a full mathematical theory of these ``numerical'' tygers is still missing in the literature,  we have addressed the problem from an entirely different perspective. 
Specifically, by focusing on the inviscid Burgers equation, we have investigated in detail the loss of time-analyticity of the velocity in Eulerian coordinates. Complex-time singularities can trigger the birth of tygers at significantly earlier times than the pre-shock, even if those singularities are far off the real-time axis.

We have developed a novel Lagrangian singularity theory, in which the pre-shock---a {\it localized} complex-time singularity in Lagrangian coordinates---leads to an {\it extended} singular landscape in Eulerian coordinates, thereby opening up the eye of the tyger (see e.g.\ Fig.\,\ref{fig:eye-tyger-twomode}).
Put differently, early-time tygers are the smoking gun for the later impending pre-shock singularity in the inviscid Burgers equation.
How these findings translate to other (inviscid) fluid flows remains to be investigated, but we speculate that, by a similar mechanism through complexified time and space, complex-time singularities in Lagrangian coordinates may be transferred to complex-time singularities in Eulerian coordinates, and vice versa.

We have analyzed early-time tygers by means of a Taylor-series representation for the velocity. For trigonometric initial conditions, the Taylor-series truncations are band-limited in Fourier space and thus come naturally with a Galerkin truncation (section~\ref{sec:pheno-asy}). 
All Fourier coefficients can be determined explicitly, which allows us to investigate tygers in a highly controlled setup.

Tygers are triggered by truncation-generated waves (e.g.\ \cite{2011PhRvE..84a6301R}), and occur at spatial locations where temporal convergence of the Taylor series is lost. We have investigated two distinct methods that strongly suppress the growth of tygers, at least within the validity regime of the present time-Taylor series approach.
The first method is tyger purging and is an adapted strategy known in the literature~\cite{2020PhRvR...2c3202M}, while the second performs a UV completion in an iterative manner.

The idea of tyger purging is to set to zero all Fourier modes within a narrow band below the Galerkin (Taylor) truncation, thereby suppressing the onset of the truncation-generated wave. Indeed, purging works well for halting early-time tygers  (section~\ref{sec:purge}). However, similarly as with the original method, some precision is lost due to the radical removal of all Fourier modes beyond a certain threshold. The loss of precision can be compensated by employing (effectively) a higher Galerkin truncation in the approaches (cf.\ left panel of Fig.\,\ref{fig:purging}). This, however, appears to be not optimal computationally, neither in the Taylor-series approach nor in a numerical setup. See also Ref.~\cite{PhysRevE.87.033017} for an interesting starting point, providing a more surgical removal of tygers using wavelets.

The second tested and novel method for reducing the growth of tygers is an iterative procedure, which successively adds more UV modes to the velocity (section~\ref{sec:UVcomplete}). 
The method requires an input solution, in the present case a low-order Taylor truncation of the velocity, and, roughly speaking, doubles  the number of Fourier modes during each iteration. Each iteration involves taking a space derivative and a temporal integration and thus, the additional UV modes come at little computational overhead. 
With each iteration, the amplitude of the tygers shrinks, and energy conservation is iteratively restored (Tabs.\,\ref{tab:tyger} and~\ref{tab:energy} respectively).

However, many questions about the UV completion do remain: 
First and foremost, a rigorous theoretical understanding of the underlying mechanism is still missing, and should in particular provide details on how exactly the energy conservation is restored. Second, the iterative method depends on the (Galerkin/Taylor) truncation of the input solution, but the present choice of (Taylor) truncation was chosen rather by heuristic means. Third, the current implementation of the UV completion is done in an iterative manner, but we speculate that there might be a non-perturbative resummation of the method (cf.\ Dyson series in quantum mechanics).

The UV completion method could offer entirely new avenues for numerical simulation techniques of general multi-dimensional fluids. It is of particular interest to relate this method to accurate subgrid-scale modelling.
For this note that the input for the iterative procedure does not need to be a truncated Taylor-series, but could be obtained from e.g.\ a simulation at a coarser spatial resolution. Also, shocks may be handled, provided one employs a weak formulation of the iterative procedure.

There are several interesting  avenues one could pursue.  
One would be to apply our methods to other fluids in higher dimensions, such as incompressible Euler flow. Questions of blow-ups could be handled by a suitably altered strategy of the Lagrangian singularity theory, and/or with an asymptotic analysis using the Mercer--Roberts extrapolation technique. For the latter, one can imagine also hybrid approaches, where one evolves the velocity with a high-resolution simulation up to some critical time, and then use this evolved velocity as the input for a local Taylor expansion.

One could also straightforwardly apply our methods to cosmological fluid flow, governed by the cosmological Euler--Poisson equations \cite{Bernardeau2002,2014JFM...749..404Z}. 
In fact, it was precisely the cosmological case that triggered the emergence of the present paper. 
Just to highlight a specific problem, it is known that time-Taylor solutions of the Euler--Poisson equations in Eulerian coordinates diverge well before the appearance of real singularities, which hinders the cosmological community to provide reliable predictions for the two-point correlation function of the matter density (e.g.\ \cite{2008PhRvD..77f3530M,2016JCAP...01..043M}).
In one-space dimension and until the pre-shock, the velocities of inviscid Burgers' and of the cosmological Euler--Poisson equations coincide exactly. Thus, the present findings  translate directly to the cosmological case.
Beyond 1D, which is of course the physically relevant case, this coincidence ceases to be true, essentially due to the presence of non-trivial gravitational interactions. Nonetheless, there exist by now various algorithms that can incorporate the gravitational interactions efficiently; to a good approximation even after the  appearance of the first singularities~\cite{2018PhRvL.121x1302S,2021MNRAS.505L..90R}; see e.g.\ \cite{2021RvMPP...5...10R} for a review.

Finally, related to tyger purging, it appears to us that this method could have strong ties to renormalization(-group) methods of general fluid flow (see e.g.\ \cite{ZHOU20101}), which should be investigated further.

\begin{acknowledgments}
We thank  Nicolas Besse, Sergey Nazarenko and Samriddhi Ray for useful discussions. O.H.\ acknowledges funding from the European Research Council (ERC) under the European Union's Horizon 2020 research and innovation programme, Grant Agreement No.\ 679145 (COSMO-SIMS).
\end{acknowledgments}

\bibliography{biblio.bib}

\end{document}